\documentclass[lettersize,journal]{IEEEtran}

\usepackage[utf8]{inputenc} 
\usepackage[T1]{fontenc}    
\usepackage{hyperref}       
\usepackage{url}            
\usepackage{booktabs}       
\usepackage{amsfonts}       
\usepackage{nicefrac}       
\usepackage{microtype}      
\usepackage{lipsum}
\usepackage{graphicx}
\usepackage{subfigure}
\graphicspath{ {./images/} }
\usepackage{amsmath,amssymb,amsthm}
\usepackage{algorithm}
\usepackage{algorithmicx}
\usepackage{algpseudocode}
\usepackage{braket}
\usepackage{booktabs}
\usepackage{threeparttable}
\usepackage{xcolor}
\usepackage{supertabular}
\usepackage{longtable}
\usepackage{multirow}
\usepackage{array}

\usepackage{tikz}
\definecolor{custompurple}{rgb}{0.5, 0.1, 0.5}
\usepackage{circuitikz}
\usetikzlibrary{quantikz}
\usepackage{setspace} 
\usepackage{circledsteps}

\pgfkeys{/csteps/inner ysep=10pt}
\usepackage{yquant}
\useyquantlanguage{groups}

\newtheorem{thm}{Theorem}
\newtheorem{lem}{Lemma}
\newtheorem{fact}{Fact}

\newtheorem{cor}{Corollary}
\newtheorem{defn}{Definition}
\newtheorem{exam}{Example}
\newtheorem{remark}{Remark}

\newcommand\vecc[1]{|#1\rangle\!\rangle}

\newcommand{\bphi}{{\Phi}}

\newcommand {\LimTDD} {{\mathit{LimTDD}}}

\newcommand {\F} {{\mathcal{F}}}
\newcommand {\G} {{\mathcal{H}}}

\newcommand {\cont} {{\texttt{cont}}}
\newcommand {\add} {{\texttt{add}}}

\newcommand {\low} {{\texttt{low}}}
\newcommand {\high} {{\texttt{high}}}
\newcommand {\locnorm} {{\texttt{loc\_norm}}}
\newcommand {\idx} {{\texttt{idx}}}
\newcommand {\w} {{\texttt{wt}}}

\newcommand {\stab} {{\texttt{Stab}}}

\usepackage[normalem]{ulem}

\newcommand{\blue}[1]{{\color{black} #1}}
\newcommand{\addt}[1]{{\color{black} #1}}

\begin{document}

\title{LimTDD: A Compact Decision Diagram Integrating Tensor and Local Invertible Map Representations}

\author{
Xin Hong$^1$, Aochu Dai$^2$, Dingchao Gao$^1$, Sanjiang Li$^3$, Zhengfeng Ji$^{2,4}$, Mingsheng Ying$^3$\\
~\IEEEmembership{$^1$Key Laboratory of System Software (Chinese Academy of Sciences) and State Key Laboratory of Computer Science, Institute of Software, Chinese Academy of Sciences, China}\\
~\IEEEmembership{$^2$Department of Computer Science and Technology, Tsinghua University, China}\\
~\IEEEmembership{$^3$Centre for Quantum Software and Information, University of Technology Sydney, Australia}\\
~\IEEEmembership{$^4$ Zhongguancun Laboratory, China}\\
\thanks{Emails: \{hongxin, gaodc\}@ios.ac.cn, dac22@mails.tsinghua.edu.cn, jizhengfeng@tsinghua.edu.cn, \{sanjiang.li, mingsheng.ying\}@uts.edu.au}

\thanks{Work partially supported by the National Natural Science Foundation of China (Grant No. 12471437, 12347104), Beijing Nova Program (Grant No. 20220484128), National Key Research and Development Program of China (Grant No.\ 2023YFA1009403), and Beijing Natural Science Foundation (Grant No.\ Z220002).
}
}



\maketitle



\begin{abstract}

\addt{Tensor networks serve as a powerful tool for efficiently representing and manipulating high-dimensional data in applications such as quantum physics, machine learning, and data compression.} Tensor Decision Diagrams (TDDs) offer an efficient framework for tensor representation by leveraging decision diagram techniques. However, the current implementation of TDDs and other decision diagrams fail to exploit tensor isomorphisms, limiting their compression potential. This paper introduces Local Invertible Map Tensor Decision Diagrams (LimTDDs), an extension of TDDs that incorporates local invertible maps (LIMs) to achieve more compact representations. Unlike LIMDD, which uses Pauli operators for quantum states, LimTDD employs the $XP$-stabilizer group, enabling broader applicability across tensor-based tasks. We present efficient algorithms for normalization, slicing, addition, and contraction, critical for tensor network applications. Theoretical analysis demonstrates that LimTDDs achieve greater compactness than TDDs and, in best-case scenarios and for quantum state representations, offer exponential compression advantages over both TDDs and LIMDDs. Experimental results in quantum circuit tensor computation and simulation confirm LimTDD's superior efficiency. \addt{Open-source code is available at \url{https://github.com/Veriqc/LimTDD}.}

\end{abstract}


\section{Introduction}

\IEEEPARstart{O}{ver} the past decade, tensor networks have evolved from a niche technique in condensed-matter physics into a universal toolkit for high-dimensional data. This transformation is driven by their ability to decompose exponentially large tensors into networks of low-rank cores, enabling high-efficiency algorithms for problems such as quantum many-body simulations, previously deemed intractable. Key milestones include DMRG-based PEPS simulators for quantum entanglement \cite{ran2020quantum}, MPS-driven discoveries of exotic quantum phases \cite{haller2023quantum}, and tensorized neural networks, which leverage tensor decomposition to match the performance of transformers on moderate-scale language-modeling benchmarks while using orders of magnitude fewer parameters \cite{ma2019tensorized}. Recent advances in automatic differentiation and GPU-accelerated contraction libraries (e.g., ITensor \cite{fishman2022itensor} and TensorNetwork-TensorFlow \cite{ganahl2019tensornetwork}) enable efficient optimization within deep-learning pipelines.



\addt{In recent years, significant progress has been made in developing compact data structures for tensor representation and tensor network computations. One such structure, the Tensor Decision Diagram (TDD), combines tensor network techniques with decision diagram methodologies to enable efficient representation and manipulation of tensor structures. Besides quantum state simulation and tensor computation,} TDDs have been successfully applied to approximate equivalence checking of noisy quantum circuits~\cite{hong2021approximate} and equivalence checking of dynamic quantum circuits~\cite{10069104}. Notably, optimizing the contraction order of tensor networks can further enhance TDDs' computational efficiency~\cite{ftdd10528748}.

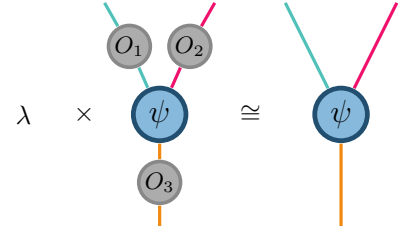
\begin{figure}[!t]
    \centering
\usetikzlibrary{backgrounds,fit,calc}
\definecolor{tensorBlue}{HTML}{224f70}
\definecolor{tensorBlueLight}{HTML}{87B9DD}
\definecolor{nodeGray}{HTML}{888888}
\definecolor{nodeGrayLight}{HTML}{ACACAC}
\definecolor{edgeOrange}{HTML}{f18810}
\definecolor{edgeMagenta}{HTML}{f0116d}
\definecolor{edgeGreen}{HTML}{50c1b8}


\begin{tikzpicture}[
  tensor ring/.style args={#1/#2}{
    circle, fill=#1, minimum size=4.8ex, inner sep=0pt,
    append after command={
      \pgfextra{\let\mytikzlastnode\tikzlastnode}
      node[circle, fill=#2, minimum size=4ex, inner sep=0pt] at (\mytikzlastnode) {}
    }
  },
  small gray ring/.style={
    circle, fill=nodeGray, minimum size=3.8ex, inner sep=0pt,
    append after command={
      \pgfextra{\let\mytikzlastnode\tikzlastnode}
      node[circle, fill=nodeGrayLight, minimum size=3.2ex, inner sep=0pt] at (\mytikzlastnode) {}
    }
  },
  tensor text/.style={inner sep=0pt, font=\large},
  small text/.style={inner sep=0pt, font=\small}
]
  \begin{scope}
    \node at (0,0) {$\lambda$};
  \end{scope}

  \begin{scope}[shift={(0.8,0)}]
    \node at (0,0) {$\times$};
  \end{scope}

  \begin{scope}[shift={(1.8,0)}]
    \node[tensor ring=tensorBlue/tensorBlueLight] (psi) at (0,0) {};
    \node[small gray ring] (O1) at (-0.4,0.9) {};
    \node[] (O1End) at (-0.8,1.6) {};
    \node[small gray ring] (O2) at (0.4,0.9) {};
    \node[] (O2End) at (0.8,1.6) {};
    \node[small gray ring] (O3) at (0,-0.9) {};
    \node[] (O3End) at (0,-1.6) {};
    
    \node[tensor text] at (psi) {$\psi$};
    \node[small text] at (O1) {$O_1$};
    \node[small text] at (O2) {$O_2$};
    \node[small text] at (O3) {$O_3$};
    
    \draw[very thick,edgeGreen] (psi) -- (O1);
    \draw[very thick,edgeMagenta] (psi) -- (O2);
    \draw[very thick,edgeOrange] (psi) -- (O3);
    \draw[very thick,edgeGreen] (O1End) -- (O1);
    \draw[very thick,edgeMagenta] (O2End) -- (O2);
    \draw[very thick,edgeOrange] (O3End) -- (O3);
  \end{scope}

  \begin{scope}[shift={(3,0)}]
    \node at (0,0) {$\cong $};
  \end{scope}
  
  \begin{scope}[shift={(4.2,0)}]
    \node[tensor ring=tensorBlue/tensorBlueLight] (psi) at (0,0) {};
    \node[] (O1End) at (-0.8,1.6) {};
    \node[] (O2End) at (0.8,1.6) {};
    \node[] (O3End) at (0,-1.6) {};

    \node[tensor text] at (psi) {$\psi$};
    
    \draw[very thick,edgeGreen] (psi) -- (O1End);
    \draw[very thick,edgeMagenta] (psi) -- (O2End);
    \draw[very thick,edgeOrange] (psi) -- (O3End);

  \end{scope}

\end{tikzpicture}

\caption{Tensor isomorphism, where $\lambda\neq 0$ is a complex number and  each $O_i$ is an invertible operator. The same edge color corresponds to the same index. \addt{After applying a local operator, we rename the corresponding index to keep it consistent with the original one.}}
\label{fig:tensoriso}
\end{figure}

While TDDs offer a compact and structured representation, they do not explicitly exploit isomorphic structures within tensors. In this paper, we define two tensors as \emph{isomorphic} if they encode the same essential information, up to a nonzero global coefficient and a local transformation applied to each index (see Fig.~\ref{fig:tensoriso} for an illustration). Exploiting these isomorphic relationships enables significant compression by merging structurally similar tensors into a single representation. 


\addt{The Local Invertible Map Decision Diagram (LIMDD), introduced in~\cite{vinkhuijzen2023limdd}, leverages a specific tensor isomorphism (i.e., quantum state isomorphism) to represent and manipulate quantum states. By identifying isomorphic relations and extracting local Pauli operators, LIMDD achieves superior compression efficiency, representing any stabilizer state with at most $n$ non-terminal nodes, where $n$ is the number of qubits. Although LIMDD has been extended to represent general tensors, it lacks mechanisms for tensor contraction and efficient tensor computations in its current implementation~\cite{vinkhuijzen2023limdd,vinkhuijzen2023efficient}.}



This paper aims to incorporate isomorphic tensor compression into a decision diagram framework. To achieve this, we propose integrating Local Invertible Maps (LIMs)~\cite{vinkhuijzen2023limdd} into TDDs, creating a versatile decision diagram that applies to tensor network tasks while maintaining at least the same compression efficiency as TDD. However, integrating LIMs into TDD presents several key challenges: 
\begin{enumerate}
\item \textbf{Efficiently identifying and utilizing tensor isomorphisms}:  While the concept of tensor isomorphism is straightforward to define (Fig.~\ref{fig:tensoriso}), recognizing and applying these transformations within a decision diagram remains nontrivial. A structured approach is required to detect and exploit isomorphic structures while ensuring TDD’s operational efficiency is maintained. 
\item \textbf{Extending beyond Pauli-based equivalences}: LIMDDs rely on Pauli operators to establish equivalences for quantum states, which limits their effectiveness to represent general tensors. A more expressive operator group is needed to enhance compression efficiency and extend applicability beyond simulating quantum states. 
\item \textbf{Ensuring a canonical and compact representation}: Ensuring a canonical and compact representation demands a systematic approach to normalizing tensor structures, which involves refining weight normalization techniques.
\item \addt{\textbf{Efficiently implementing the contraction operation}: Efficient tensor network contraction operations require making full use of the isomorphism relationships among tensors. This requires us to handle the local operators on the edges of the decision diagram properly.
}

\end{enumerate}

To address these challenges, we introduce the Local Invertible Map Tensor Decision Diagram (LimTDD), an extension of TDD that incorporates LIMs for enhanced compression. Our main contributions include:
\begin{enumerate}
    \item Addressing Challenge 1, we formalize tensor vectorization and present procedures to identify tensor isomorphisms within the LimTDD structure, enabling more effective compression while maintaining computational efficiency.
    \item Addressing Challenge 2, we utilize the $XP$-stabilizer group \cite{webster_xp_2022} in LimTDD, extending its applicability beyond quantum states and improving compression efficiency compared to Pauli-based LIMDD. 
    \item Addressing Challenges 3, we develop optimized algorithms for normalization and provide theoretical proofs to ensure the canonicity and compactness of LimTDD.
    \item \addt{Addressing Challenges 4, we enhance the computational efficiency of LimTDD by removing the operators applied to open indices and transfer operators applied to contracted indices to one LimTDD, thus reducing the complexity of the LimTDDs that need to be contracted and improving the efficiency of reusing intermediate results.
    } 
    \item Additionally, we conduct extensive experiments demonstrating its advantages in compactness and computational efficiency for quantum circuit simulation and functionality computation.
\end{enumerate}

\subsection*{Related Work}

\addt{
Decision diagrams, originally developed for analyzing Boolean functions, have been widely adopted across various fields through their diverse variants. By incorporating multi-terminal nodes~\cite{fujita1997multi} or valued edges~\cite{vrudhula1996edge}, decision diagrams have been extended to applications such as pseudo-Boolean functions~\cite{vrudhula1996edge,tafertshofer1997factored}, probabilistic distributions~\cite{kisa2014probabilistic}, algebraic systems~\cite{bahar1997algebric, sanner2005affine, wilson2005decision}, and matrix operations~\cite{vrudhula1996edge,tafertshofer1997factored}.
}

\addt{
To meet the demands of quantum system simulation and verification, several decision diagram models supporting complex domains and matrix manipulations have been introduced, including Quantum Information Decision Diagrams (QuIDD)~\cite{viamontes2003improving}, Quantum Multiple-Valued Decision Diagrams (QMDD)~\cite{niemann2015qmdds}, X-Decomposition Quantum Decision Diagrams (XQDD)~\cite{wang2008xqdd}, and Context-Free-Language Ordered Binary Decision Diagrams (CFLOBDD)~\cite{sistla2024cflobdds}. These decision diagrams have been successfully applied to quantum circuit simulation and verification~\cite{viamontes2004high,burgholzer2020advanced,burgholzer2021qcec}, quantum state preparation~\cite{DBLP:conf/dac/MatoHW24}, and quantum circuit design automation~\cite{DBLP:conf/qsw/WilleBFKMPQRSSSSB24}. Specifically, QuIDD adapts classical algebraic decision diagrams~\cite{bahar1997algebric} for quantum information, QMDD leverages the unique structure of quantum circuits, XQDD exploits symmetries in quantum gates, and CFLOBDD organizes diagram structures using a branching program and procedure-calling approach. Additionally, knowledge maps for decision diagrams have been explored to enhance their representational capabilities~\cite{fargier2014knowledge,vinkhuijzen2024knowledge}.
}

The remainder of this paper is structured as follows. Section~\ref{sec:background} provides background on quantum computing, tensor networks, and decision diagrams (TDD and LIMDD). Section~\ref{sec:LimTDD} introduces the formal definition of LimTDD, along with its mathematical foundations. Section~\ref{sec:Algorithms} describes key tensor operations for LimTDD, including slicing, addition, and contraction. Section~\ref{sec:compare} presents a theoretical comparison between LimTDD, TDD, and LIMDD, analyzing their compression efficiency. 
Section~\ref{sec:experiments} provides experimental results, demonstrating LimTDD’s advantages in quantum circuit simulation and tensor network computations. Finally, Section~\ref{sec:conclusion} concludes the paper. For clarity, proofs are provided in the appendix.

\section{Background}\label{sec:background}

\subsection{Quantum Computing}

Quantum computing leverages principles of quantum mechanics to perform computations that classical computers cannot efficiently handle. The fundamental unit of quantum information is the \textit{qubit}, which differs from a classical bit by existing in a superposition of states. A single-qubit state is generally represented as:
\begin{equation}
\label{eq:qubit}
  \ket{ \varphi} :=  \alpha_0 \ket{0} + \alpha_1 \ket{1},
\end{equation}
where $\alpha_0$, $\alpha_1$ are complex amplitudes satisfying ${\left| \alpha_0  \right|^2} + {\left| \alpha_1  \right|^2} = 1$. Alternatively, a single-qubit state can be represented as a column vector $[\alpha_0, \alpha_1]^\intercal$ (where $^\intercal$ denotes the transpose). In general, an $n$-qubit quantum state can be expressed as a $2^n$-dimensional vector $[\alpha_0, \alpha_1, \dots, \alpha_{2^n-1}]^\intercal$ in Hilbert space.



Quantum computation is performed using quantum gates, which are represented by unitary matrices acting on qubits. Common quantum gates include the Pauli gates ($X$, $Y$, $Z$), the Hadamard gate ($H$), and controlled operations such as the Controlled $Z$ ($CZ$) gate and the Controlled Not (CNOT) gate (cf.~\cite{nielsen2010quantum}). Each quantum gate has a unique unitary matrix representation in a predefined orthonormal basis.
Of particular importance to this paper is the Phase gate
$$P = \left[\begin{array}{cccc} 
		1 & 0\\ 
		0 & \omega^2\\
\end{array}\right],
$$
which is defined with a parameter $\omega \equiv e^{\pi i/N}$, the $2N$-th root of the unity. To emphasize, we also denote $P$ as $P_N$. In this paper, if not specified, $N=8$, $\omega = e^{\pi i/8}$, and $P, P^2, P^4$ correspond to gates $T, S, Z$, respectively.

The outcome of applying a quantum gate to an input state is determined by multiplying the corresponding unitary matrix with the vector representing the input quantum state. For instance, the result of applying an $X$ gate to the input state $\left[{\alpha _0}, {\alpha _1}\right]^\intercal$ is given by $\left[{\alpha _1}, {\alpha _0}\right]^\intercal$. In a broader context, an $n$-qubit quantum gate is represented by a $2^n\times 2^n$-dimensional unitary transformation matrix.

A quantum circuit is composed of qubits and a sequence of quantum gates. When a quantum circuit operates on an input state, the quantum gates are applied sequentially. The functionality of an $n$-qubit quantum circuit is described by a $2^n\times 2^n$-dimensional unitary transformation matrix.

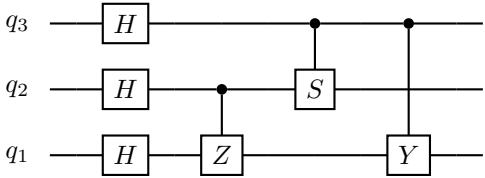
\begin{figure}[h]
\centerline{
\begin{tikzcd}[column sep=0.35cm,row sep=0.35cm]
\lstick{$q_3\ $}&\qw &\gate{H} &\qw &\qw      &\qw  &\ctrl{1}  &\qw  &\ctrl{2} &\qw&\qw\\
\lstick{$q_2\ $}&\qw &\gate{H} &\qw &\ctrl{1} &\qw  &\gate{S}  &\qw  &\qw      &\qw&\qw\\
\lstick{$q_1\ $}&\qw &\gate{H} &\qw &\gate{Z} &\qw  &\qw       &\qw  &\gate{Y} &\qw&\qw\\
\end{tikzcd}

}
    \caption{A quantum circuit with 3 qubits and 6 gates.}
    \label{fig:qc}
\end{figure}

\begin{exam}\label{exp:qc}
Consider the circuit shown in Fig.~\ref{fig:qc}. This circuit consists of 3 Hadamard gates, a controlled-$Z$ gate, a controlled-$S$ gate, and a controlled-$Y$ gate. Starting from the input state $\ket{000}$, the output state of this circuit is $\frac{1}{2\sqrt{2}}(\ket{000}+ \ket{001}+\ket{010}-\ket{011}-i\ket{100}+i\ket{101}-\ket{110}-\ket{111})$.
\end{exam}

\subsection{Tensors and Tensor Networks}

Tensor networks are a powerful framework for representing and manipulating high-dimensional data. They have become indispensable in areas such as quantum circuit simulation, quantum many-body physics, and quantum chemistry, where they enable efficient data compression and scalable computation.
Beyond physics, tensor-network techniques have also been applied to 
machine learning and data modeling, offering both computational efficiency and physical interpretability.


A \emph{tensor} can be formally defined as a mapping from a set of indices to the complex numbers. In quantum computing, each index typically takes values in $\{0,1\}$. Given an index set $S = \{x_n, \ldots, x_1\}$, a tensor is a function $\phi : \{0,1\}^S \rightarrow \mathbb{C}$, where $\mathbb{C}$ denotes the field of complex numbers. For clarity, we denote such a tensor as $\phi_{x_n \ldots x_1}$ or $\phi_{\vec{x}}$, 
and its value under the assignment $\{x_i \mapsto a_i \mid 1 \le i \le n\}$ as 
$\phi_{x_n \ldots x_1}(a_n, \ldots, a_1)$, or simply $\phi_{\vec{x}}(\vec{a})$ or $\phi(\vec{a})$ when unambiguous. 
The \emph{rank} of a tensor is the number $n$ of its indices. 
Scalars, two-dimensional vectors, and $2\times 2$ matrices correspond to rank-0, rank-1, and rank-2 tensors, respectively. 
An $n$-qubit quantum state can be regarded as a rank-$n$ tensor.

The primary operation between tensors is \emph{contraction}. Given tensors $\gamma_{\vec{x}, \vec{z}}$ and $\xi_{\vec{y}, \vec{z}}$ sharing  a common index set $\vec{z}$, their contraction results in a new tensor $\phi_{\vec{x}, \vec{y}}$ defined as:
\begin{equation}\label{eq:contdef}
	\phi_{\vec{x}, \vec{y}}(\vec{a}, \vec{b})=\sum_{\vec{c}\in \{0,1\}^{\vec{z}}}{\gamma_{\vec{x}, \vec{z}}(\vec{a}, \vec{c})\cdot \xi_{\vec{y}, \vec{z}}}(\vec{b}, \vec{c}).
\end{equation}

Another useful operation is \emph{slicing}, analogous to the cofactor operation for Boolean functions. For a tensor $\phi$ with index set $S=\{x, x_n,\ldots,x_1\}$, the slicing with respect to $x = c$ (where $c\in \{0,1\}$) is a tensor $\phi_{x=c}$ over $S'=\{x_n, \ldots, x_1\}$ given by:
\begin{align}
\phi_{x=c}(\vec{a}):= \phi(c, \vec{a})
\end{align}
for any $\vec{a} \in \{0,1\}^n$. The \textit{negative} and \textit{positive} slicings of $\phi$ with respect to $x$ are denoted as $\phi_{x=0}$ and $\phi_{x=1}$, respectively. Suppose $\phi$ does not rely on $x$; that is $x$ is not in the index set of $\phi$. We denote $\phi_{x=0}=\phi_{x=1}=\phi$.

A \emph{tensor network} is represented by an undirected graph $G=(V, E)$ with zero or multiple open edges, where each vertex $v$ in $V$ represents a tensor, and each edge corresponds to a shared index between two adjacent tensors. By contracting connected tensors (i.e., vertices in $V$) in an arbitrary order, we obtain a rank $m$ tensor, where $m$ is the number of open edges of $G$. This tensor, which is independent of the contraction order, is called the tensor representation of the tensor network. For further details, see~\cite{markov2008simulating} and~\cite{biamonte2019lectures}.


Quantum circuits provide natural examples of tensor networks. In this paper, we use quantum-circuit applications as the primary examples to illustrate our framework.

\begin{exam}\label{exp:tensor}
 The circuit shown in Fig.~\ref{fig:qc} can be represented as a tensor network consisting of three rank-2 tensors and three rank-4 tensors. Contracting these tensors with a rank-3 tensor representing the input state $\ket{000}$ yields a rank-3 tensor $\phi_{x_3x_2x_1}$ as follows, which represents the output state as shown in Example~\ref{exp:qc}. Here, we use $x_3,\ x_2,\ x_1$ to represent the indices corresponding to the output of qubits $q_3,\ q_2,\ q_1$.


\begin{table}[H]
    \centering
    \setlength{\tabcolsep}{4pt} 
    \renewcommand\arraystretch{1.6} 
    \begin{tabular}{c|cccccccc}
        \( x_3x_2x_1 \) & 000 & 001 & 010 & 011 & 100 & 101 & 110 & 111 \\ \hline
        \( \phi \) & \(\frac{1}{2\sqrt{2}}\) & \(\frac{1}{2\sqrt{2}}\) & \(\frac{1}{2\sqrt{2}}\) & \(\frac{-1}{2\sqrt{2}}\) & \(\frac{-i}{2\sqrt{2}}\) & \(\frac{i}{2\sqrt{2}}\) & \(\frac{-1}{2\sqrt{2}}\) & \(\frac{-1}{2\sqrt{2}}\) \\ 
    \end{tabular}
    \label{tab:tensor}
\end{table}

    
         

\end{exam}


\subsection{Local Invertible Maps and $XP$-Operators}\label{sec:xplim}
\begin{defn}[Local Invertible Maps \cite{vinkhuijzen2023limdd}]\label{def:lim}


An $n$-qubit Local Invertible Map (LIM) is an operator $O$ of the form 
\begin{equation}\label{eq:limO}
    O = \lambda O_n \otimes \cdots \otimes O_1, 
\end{equation} 
where each $O_i$ is an invertible $2\times 2$ matrix and $\lambda \in \mathbb{C}$. The set of all such maps is denoted as $\mathcal{O}_n$, and the set of all LIMs is defined as 
\begin{equation}\label{eq:limCalO}
\mathcal{O} = \bigcup_{n\in \mathbb{N}}{\mathcal{O}_n}.
\end{equation}   
\end{defn}

Let $O=\lambda \cdot O_n\otimes \cdots \otimes O_1$ be an LIM. We write $c(O)=\lambda$.
Suppose $\lambda = r\cdot e^{2\pi i \theta}$ for some real number $r \in \mathbb{R}$. We call $\theta$ the angle of $\lambda$, denoted as  $\texttt{ang}(\lambda)=\theta$.

In practice, the operators are restricted to a smaller group for computational efficiency. Let $G = \{g_1,\cdots, g_m\}$ be a set of LIMs. We denote the group generated by $G$ as $\langle G\rangle$.

Let $\mathcal{G}$ be a subgroup of $\mathcal{O}_n$. For an $n$-qubit quantum state $\ket{\psi}$, an operator $g\in \mathcal{G}$ is a $\mathcal{G}$-stabilizer of $\ket{\psi}$ if $g\ket{\psi}=\ket{\psi}$.  The $\mathcal{G}$-stabilizer group of $\ket{\psi}$ is defined as
\begin{equation}\label{eq:stab_state}
\stab(\ket{\psi}) = \{g\in \mathcal{G}\ |\ g\ket{\psi} =\ket{\psi}\}.  
\end{equation}



An $n$-qubit LIM $O=\lambda O_n \otimes \cdots \otimes O_1$ is a Pauli operator if $\lambda\in \set{1,-1,i,-i}$ and each $O_i$ is $I$ or a Pauli gate $X,Y,Z$. 

$XP$-operators are more general than Pauli operators. An $n$-qubit $XP$-operator of precision $N$ (where $\omega=e^{\pi i/N}$) is an operator of the form:
\begin{equation}
   XP_N(p|\bold{x}|\bold{z}) := \omega^p \bigotimes_{0\leq i<n}X^{\bold{x}[i]}P^{\bold{z}[i]}, 
\end{equation}
where $p$, $\bold{x}[i]$, and $\bold{z}[i]$ are integers satisfying:
\begin{itemize}
    \item $0\leq p<2N$, 
    \item $0\leq \bold{x}[i]<2$, 
    \item $0\leq \bold{z}[i]<N$.
\end{itemize}
We call $p$, $\bold{x}$, and $\bold{z}$ the phase component, $X$-component, and $Z$-component of the $XP$-operator, respectively. It is easy to see that two $XP$-operators $XP_N(p|\bold{x}|\bold{z})$ and $XP_N(p'|\bold{x'}|\bold{z'})$ are identical if and only if $p=p'$, $\bold{x}=\bold{x'}$, and $\bold{z}=\bold{z'}$.

As established in \cite[Proposition 4.1]{webster_xp_2022}, $XP$-operators possess useful group properties. Examples include elegant characterizations of the unit operator, multiplication, and inverses of $XP$-operators. Of particular importance is the following result: Let $G = \{g_1,\cdots, g_m\}$ be a set of $XP$-operators. There exists a unique set of diagonal operators $S_Z:=\{ B_j:0 \leq j <s\}$ and non-diagonal operators $ S_X:=\{ A_i:0\leq i<r\}$, called the \emph{canonical generators} of $\langle G\rangle$, such that all group elements $g \in \langle G\rangle$ can be expressed as $g=S_X^{\bold{a}} S_Z^{\bold{b}}$, where $\bold{a} \in \mathbb{Z}_2^{|S_X|}$ and $\bold{b} \in \mathbb{Z}_2^{|S_Z|}$. Furthermore, two sets of $XP$-operators of precision $N$ generate the same group if and only if they have the same canonical generators. 

We denote by \(\mathcal{XP}_N\) the group generated by \(XP\)-operators with precision \( N \). When the precision \( N \) is not specified, we use the shorthand \(\mathcal{XP}\).


\subsection{LIMDD for Quantum State Representation}

 
Quantum circuit simulation is typically computationally intensive. LIMDD \cite{vinkhuijzen2023limdd} offers a compact representation of quantum states, enabling efficient simulation. Within this framework, quantum states that differ only by a local invertible transformation on each qubit can share the same memory space.

\begin{defn}[Quantum State Isomorphism \cite{vinkhuijzen2023limdd}]
Two $n$-qubit quantum states $\ket{\Phi}$ and $\ket{\Psi}$ are said to be \emph{isomorphic} if $\ket{\Phi} = O \ket{\Psi}$ for some $O \in \mathcal{O}_n$.
\end{defn}

With the notion of quantum state isomorphism in place, we now introduce Local Invertible Map Decision Diagrams.

\begin{defn}[LIMDD \cite{vinkhuijzen2023limdd}]\label{def:limdd}
Let $\mathcal{G}$ be a subgroup of $\mathcal{O}$. An $n$-qubit $\mathcal{G}$-LIMDD is a rooted, directed acyclic graph (DAG) that represents an $n$-qubit quantum state. Formally, a LIMDD is a 6-tuple $(V, E, \idx, \low, \high, \w)$ where:
    \begin{itemize}
	\item $V$ is a finite set of nodes, consisting of non-terminal nodes $V_{NT}$ and a terminal node $v_T$ labeled with integer 1;
        \item  $\idx: V_{NT} \rightarrow [n]$ assigns each node a qubit index in $[n]$;
	\item $\low$ and $\high$ are mappings in $V_{NT} \rightarrow V$ that assign each non-terminal node its 0-successor and 1-successor, respectively;
	\item $E = \{(v, \low(v)), (v, \high(v)) : v\in V_{NT}\}$ is the set of edges, where $(v, \low(v))$ and $(v, \high(v)) $ are called the low-edge and high-edge of $v$, respectively. For simplicity, we assume the root node $r_\F$ has a unique incoming edge, denoted  $e_r$, which has no source node;
        \item  $\w: E \rightarrow \mathcal{G}$ is a function that labels edges with LIMs.
\end{itemize}

By overloading the Dirac notation, the semantics of the terminal node is defined to be $\ket{v_T} = 1$, the semantics of an edge $e$, directing to a node $v$, is defined to be 
$$
\ket{e} = w(e) \cdot \ket{v},
$$
and the semantics of a non-terminal node $v$ is defined to be 
$$
\ket{v} = \ket{0}\otimes \ket{(v,\low(v))} + \ket{1}\otimes \ket{(v,\high(v))}.
$$

\end{defn}

In the LIMDD framework \cite{vinkhuijzen2023limdd} and implementation \cite{vinkhuijzen2023efficient}, the group $\mathcal{G}$ is chosen as the Pauli group, i.e., the group of Pauli operators. By leveraging isomorphisms among quantum states, LIMDDs achieve greater compactness than QMDDs and TDDs in representing quantum states.

\begin{exam}\label{exp:LIMDD}
\addt{
Fig.~\ref{fig:run_exp}\,(a) illustrates the LIMDD representation of the 3-qubit quantum state introduced in Example~\ref{exp:qc}. 
In the diagram, dotted red lines denote low edges, solid blue lines denote high edges, and the purple edge indicates the incoming edge. 
The terminal node represents the constant value~1. For clarity, edge weights equal to $1$ or $I$ are omitted there.

The node labeled $x_1$ corresponds to the state 
$\ket{+} = \ket{0} + \ket{1}$.
The left node labeled $x_2$ represents the 2-qubit state
$\ket{0}\ket{+} + \ket{1}(Z\ket{+}) 
 = \ket{0}\ket{+} + \ket{1}\ket{-}$; and the right node labeled $x_2$ represents
$\ket{0}\ket{+} + \ket{1}(iZ\ket{+}) 
 = \ket{0}\ket{+} + i\ket{1}\ket{-}$.
The node labeled $x_3$ represents the 3-qubit state
\begin{align*}
\ket{\Phi}=&\ket{0}(\ket{0}\ket{+}+\ket{1}\ket{-})
+ \ket{1}(iZ\!\otimes\! Z)(\ket{0}\ket{+}+i\ket{1}\ket{-}) \\
 = & \ket{0}\ket{0}\ket{+}+\ket{0}\ket{1}\ket{-}
 + i\ket{1}\ket{0}\ket{-}+\ket{1}\ket{1}\ket{+}.
\end{align*}
Finally, the entire LIMDD represents the normalized state
\begin{align*}
& \frac{1}{2\sqrt{2}}\,
(Z\!\otimes\! I\!\otimes\! I) \ket{\Phi} \\
= & \frac{1}{2\sqrt{2}}
(
\ket{0}\ket{0}\ket{+}
+\ket{0}\ket{1}\ket{-}
-i\ket{1}\ket{0}\ket{-}
-\ket{1}\ket{1}\ket{+}
).
\end{align*}
}
    
\end{exam}

\subsection{TDD for Tensor Network Representation}


TDDs provide a data structure to represent and manipulate tensors compactly and efficiently. Their core concept employs a binary branching structure to map data elements to distinct paths, enabling compression by sharing common branches for repeated data. The formal definition is as follows:

\begin{defn}[TDD \cite{hong2020tensor}]
	A \emph{Tensor Decision Diagram} (TDD) $\F$ over a set of indices $S$ is a rooted and weighted DAG 
	$\F = (V, E, \idx, \low, \high, \w)$ such that:
	\begin{itemize}
        \item $V$, $V_{NT}$, $v_T$, $\low$, $\high$, and $E$ are interpreted analogously to their roles in LIMDDs, representing the nodes, non-terminal nodes, the terminal node, 0- and 1-successor mappings, and edges. In particular, we write $r_\F$ and $e_r$ for the root node and the incoming edge of $\F$.
		\item $\idx: V_{NT} \rightarrow S$ assigns each non-terminal node an index in $S$;
		\item $\w: E\rightarrow \mathbb{C}$ assigns each edge a complex weight. In particular, $\w(e_r)$ is called the weight of $\F$, denoted $w_\F$.
	\end{itemize} 
\end{defn}

Each node $v$ of a TDD $\F$ corresponds to a tensor $\bphi(v)$.  
If $v=v_T$ is the terminal node, then $\bphi(v)=1$; 
otherwise, 
\begin{equation}\label{eq:tdd-intp}
\bphi(v) :=  w_0\cdot \overline{x}_v \cdot \bphi(\low(v))
+w_1 \cdot x_v \cdot \bphi(\high(v)),
\end{equation}
where $x_v=\idx(v)$, and $w_0$ and $w_1$ are the weights on the low- and high-edges of $v$, respectively. The tensor represented by $\F$ itself is defined to be 
\begin{align}\label{eq:Phi_of_TDD}
\bphi(\F) := w_\F \cdot \bphi(r_\F).
\end{align}

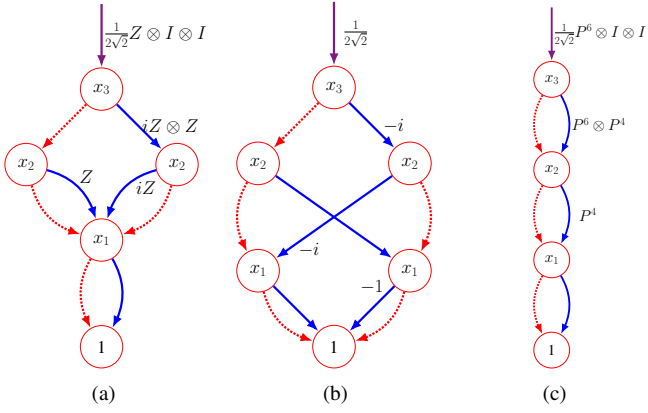
\begin{figure}
    \centering
    \subfigure[]{
        \resizebox{0.155\textwidth}{!}{


\begin{tikzpicture}[
    >=Latex, 
    line join=bevel,
    node distance=3cm, 
    every node/.style={minimum size=2cm, very thick, font=\Huge}, 
    every path/.style={line width=3pt} 
]

\node (x3) [circle, draw, red, text=black] {$x_3$};
\node (x20) [circle, draw, red, below left=of x3, text=black] {$x_2$};
\node (x21) [circle, draw, red, below right=of x3, text=black] {$x_2$};
\node (x1) [circle, draw, red, below right=of x20, text=black] {$x_1$};
\node (t) [circle, draw, red, below=of x1, text=black] {1};
\node (none) [above=of x3] {};

\draw[->, red,dotted] (x3) -- (x20) node[midway, left] {};
\draw[->, blue] (x3) -- (x21) node[midway, right,text=black] {$iZ\otimes Z$};

\draw[->, red , dotted] (x20) to[bend right] node[midway, left] {} (x1);
\draw[->, blue] (x20) to[bend left] node[near start, right,text=black] {$Z$} (x1);

\draw[->, red , dotted] (x1) to[bend right] node[midway, left] {} (t);
\draw[->, blue] (x1) to[bend left] node[midway, right] {} (t);

\draw[->, red , dotted] (x21) to[bend left] node[midway, right] {} (x1);
\draw[->, blue] (x21) to[bend right] node[midway, right,text=black] {$iZ$} (x1);

\draw[->, custompurple] (none) -- (x3) node[midway, right,text=black] {$\frac{1}{2\sqrt{2}} Z\otimes I \otimes I$};

\end{tikzpicture}

    }
    }
    \subfigure[]{%
        \resizebox{0.15\textwidth}{!}{


\begin{tikzpicture}[
  >=Latex, 
  line join=bevel,
  node distance=3cm, 
  every node/.style={minimum size=2cm, very thick, font=\Huge}, 
  every path/.style={line width=3pt} 
]

\node (x3) [circle, draw, red, text=black] {$x_3$};
\node (x20) [below left=of x3, circle, draw, red, text=black] {$x_2$};
\node (x21) [below right=of x3, circle, draw, red, text=black] {$x_2$};
\node (x10) [below=of x20, circle, draw, red, text=black] {$x_1$};
\node (x11) [below=of x21, circle, draw, red, text=black] {$x_1$};
\node (t) [below right=of x10,circle, draw, red, text=black] {1};
\node (none) [above=of x3] {};

\draw [->,red, dotted] (x3) -- (x20);
\draw [->,blue] (x3) -- (x21) node[midway,right,text=black,text=black] {$-i$};
\draw [->,red, dotted] (x20) to[bend right] (x10);
\draw [->,blue] (x20) -- (x11);
\draw [->,red, dotted] (x10) to[bend right] (t);
\draw [->,blue] (x10) -- (t);
\draw [->,red, dotted] (x11) to[bend left] (t);
\draw [->,blue] (x21) -- (x10) node[pos = 0.9, right, text=black] {$-i$};
\draw [->,red, dotted] (x21) to[bend left] (x11);
\draw[->, custompurple] (none) -- (x3) node[midway,text=black, right,text=black] {$\frac{1}{2\sqrt{2}}$};
\draw [->, blue] (x11) -- (t) node[midway, above, text=black] {$-1$};

\end{tikzpicture}

}
    }
    \subfigure[]{
        \hspace{2em}
        \resizebox{0.091\textwidth}{!}{


\begin{tikzpicture}[
  >=Latex, 
  line join=bevel,
  node distance=3cm, 
  every node/.style={minimum size=2cm, very thick, font=\Huge}, 
  every path/.style={line width=3pt} 
]

\node (x3) [circle, draw, red, text=black] {$x_3$};
\node (x2) [below=of x3, circle, draw, red, text=black] {$x_2$};
\node (x1) [below=of x2, circle, draw, red, text=black] {$x_1$};
\node (t) [below=of x1, circle, draw, red, text=black] {1};
\node (none) [above=of x3] {}; 

\draw [->, red, dotted] (x3) to[bend right] (x2);
\draw [->, blue] (x3) to[bend left] node[midway, right, text=black] {$P^{6} \otimes P^{4}$} (x2);
\draw [->, red, dotted] (x2) to[bend right] (x1);
\draw [->, blue] (x2) to[bend left] node[midway, right, text=black] {$P^{4}$} (x1);
\draw [->, red, dotted] (x1) to[bend right] (t);
\draw [->, blue] (x1) to[bend left] node[midway, right] {} (t);
\draw [->, blue, custompurple] (none) -- (x3) node[midway, right, text=black] {$\frac{1}{2\sqrt{2}} P^{6} \otimes I \otimes I$};

\end{tikzpicture}

    }
    }
    \caption{ An intuitive comparison among (a) LIMDD, (b) TDD, and (c) LimTDD. All decision diagrams represent the output state resulted from simulating the circuit shown in Fig.~\ref{fig:qc} with input state $\ket{000}$. The dotted red lines correspond to low edges and the solid blue lines correspond to high edges. The amplitude of the state can be obtained by multiplying the weights along a path. When $Z$ or $P$ is encountered, a phase $-1$ or $e^{2\pi i/8}$ should be added when going along the high edges. Here, $P = P_8$.
    }
    
    \label{fig:run_exp}
\end{figure}

\begin{exam}
Fig.~\ref{fig:run_exp}\,(b) shows the TDD of the tensor introduced in Example~\ref{exp:tensor}. 
The tensors represented by the nodes can be constructed from bottom to top, following the same procedure as in Example~\ref{exp:LIMDD}. 
The value of each tensor element is obtained by multiplying the weights along the corresponding path from the root to the terminal node. 
For instance, the leftmost path, denoted by the binary string $000$, yields
\[
\phi_{x_3x_2x_1}(000)
= \tfrac{1}{2\sqrt{2}} \times 1 \times 1 \times 1
= \tfrac{1}{2\sqrt{2}}.
\]
\end{exam}

\section{LimTDD} \label{sec:LimTDD}

\addt{While TDD enables tensor compression, it suffers from relatively lower efficiency, especially when compared with the high compressing ratio of LIMDD in representing quantum states.} In this section, we introduce LimTDD, a novel approach \addt{inspired by LIMDD that enhances TDD efficiency.}  By combining the respective advantages of TDD and LIMDD, LimTDD not only retains the ability to represent arbitrary tensors but also achieves greater compactness, offering enhanced efficiency and broader applicability.

\subsection{Isomorphism Between Tensors}\label{sec:tensoriso}

To effectively represent tensors using LimTDD, we first characterize tensor isomorphism by introducing tensor vectorization. Let $\mathcal{G}$ be a subgroup of $\mathcal{O}$. As illustrated in Fig.~\ref{fig:tensoriso}, two tensors are considered $\mathcal{G}$-isomorphic if they differ only by a global coefficient and a local invertible transformation $O\in\mathcal{G}$. In order to express this isomorphic relation precisely,
we introduce the notion of tensor vectorization:

\begin{defn}[Tensor Vectorization \cite{biamonte2019lectures}]
Let $\phi_{x_n\cdots x_1}$ be a tensor with $n$ indices. Its \addt{vector representation with respect to the index order $x_n \prec \cdots \prec x_1$}, denoted by $\vecc{\phi_{x_n\cdots x_1}}$, is a $2^n$-dimensional vector whose $i$-th element is $\phi(i_n,\cdots, i_1)$, where $i_n\cdots i_1$ is the binary expansion of the integer $i$.
\end{defn}
\addt{For instance, consider the rank-2 tensor $\phi_{xy}$ represented by  matrix $\left[\begin{array}{cccc} 
		1 & 2\\ 
		0 & 3\\
\end{array}\right],$ 
where $x$ and $y$ denote the column and row indices, respectively. Its vector representation is $\vecc{\phi_{xy}} = [1,0,2,3]^\intercal$. 
For the special case when a rank-$n$ tensor $\phi$ represents an $n$-qubit quantum state $\ket{\psi}$, then, under the same index/qubit order, $\vecc{\phi}$ coincides with the vector representation of the quantum state $\ket{\psi}$.
}



\addt{Tensor vectorization provides a convenient way to define how an operator acts on a tensor. For instance, the action of an operator $O_3 \otimes O_2 \otimes O_1$ on tensor $\phi_{x_3x_2x_1}$ can be mathematically expressed as
\[
\vecc{\phi'_{x_3x_2x_1}} = (O_3 \otimes O_2 \otimes O_1)\,\vecc{\phi_{x_3x_2x_1}}.
\]
by using the matrix-vector multiplication.

\begin{exam}\label{exp:vecc}
Consider the tensor $\phi_{x_3x_2x_1}$ given in Example~\ref{exp:tensor}.  
Its vector representation is
\[
\vecc{\phi_{x_3x_2x_1}} = \tfrac{1}{2\sqrt{2}}[1,\,1,\,1,\,-1,\,-i,\,i,\,-1,\,-1]^\intercal,
\]
which is consistent with the table representation in Example~\ref{exp:tensor} while offering greater mathematical convenience for computation.  
Suppose we apply the operator $P^2 \otimes I \otimes I = \texttt{diag}(1,1,1,1,i,i,i,i)$ to this tensor.  
The resulting tensor has the vector representation  
\[
(P^2 \otimes I \otimes I)\,\vecc{\phi_{x_3x_2x_1}}
= \tfrac{1}{2\sqrt{2}}[1,\,1,\,1,\,-1,\,1,\,-1,\,-i,\,-i]^\intercal,
\]
which corresponds to contracting the operator $P^2 = \texttt{diag}(1,i)$ with the index~$x_3$. 
\end{exam}
}

\addt{With vectorization, we could describe the isomorphism relation between tensors (as shown in Fig.~\ref{fig:tensoriso}) mathematically.} 
\begin{fact}\label{fact}
    Two rank-$n$ tensors $\phi_{x_n\cdots x_1}$ and $\psi_{x_n\cdots x_1}$ are $\mathcal{G}$-isomorphic if and only if there exist a nonzero complex number $\lambda$ and $O=\lambda \cdot O_n\otimes \cdots \otimes O_1 \in \mathcal{G}$ such that  $\vecc{\phi_{x_n\cdots x_1}} = \lambda \cdot O_n\otimes \cdots \otimes O_1 \vecc{\psi_{x_n\cdots x_1}}$.
\end{fact}

This is because, apart from a global coefficient, applying $\lambda \cdot O_n\otimes \cdots \otimes O_1$ to $ \vecc{\psi_{x_n\cdots x_1}}$ is equivalent to contracting an operator $O_i$ over each index $x_i$ of $\psi$. The key point is to ensure that the indices of both tensors remain aligned.




In the remainder of this paper, when there is no ambiguity, we will not distinguish between $\phi_{x_n\cdots x_1}$ and $\vecc{\phi_{x_n\cdots x_1}}$. Additionally, we will use the notations $\lambda \cdot \phi_{x_n\cdots x_1}$ and $O \cdot \phi_{x_n\cdots x_1}$ to represent multiplying a tensor by a scalar or contracting an operator on corresponding indices. Specially, $\vecc{O \cdot \phi_{x_n\cdots x_1}}=O \cdot \vecc{\phi_{x_n\cdots x_1}}$.


\subsection{LimTDD: Definition and Examples}


\begin{defn}[LimTDD]\label{def:limtdd}
	Let $\mathcal{G}$ be a subgroup of $\mathcal{O}$. A $\mathcal{G}$-LimTDD $\mathcal{F}$ over a set of indices $S$ is a rooted and weighted DAG 
	$\mathcal{F} = (V, E, \idx, \low, \high, \w)$ defined as follows:
	\begin{itemize}
        \item $V$, $V_{NT}$, $v_T$, $\low$, $\high$, and $E$ are interpreted analogously to their roles in LIMDDs, representing the nodes, non-terminal nodes, the terminal node, 0- and 1-successor mappings, and edges. In particular, we write $r_\F$ and $e_r$ for the root node and the incoming edge of $\F$.
        \item $\idx: V_{NT} \rightarrow S$ assigns each non-terminal node an index in $S$. We call $\idx(r_\mathcal{F})$ the top index of $\F$, if $r_\mathcal{F}$ is not the terminal node;
		\item $\w: E\rightarrow \mathcal{G}$ assigns each edge a weight in $\mathcal{G}$.  $\w(e_r)$ is called the weight of $\mathcal{F}$, and denoted $w_\mathcal{F}$.  
	\end{itemize} 

\end{defn}



\addt{As in LIMDD and TDD, the terminal node $v_T$ represents the scalar~1. For any edge $e$ pointing to a node $v'$ that represents a rank-$k$ tensor, 
the tensor associated with $e$ is also a rank-$k$ tensor  satisfies 
\begin{align}\label{eq:vecc-edge}
\Phi(e) = \w(e) \cdot \Phi(v').
\end{align}
For any internal node $v$ with index $x_v$, suppose both of its outgoing edges 
represent rank-$k$ tensors. Then $v$ itself represents a rank-$(k\!+\!1)$ tensor $\Phi(v)$, 
whose two slices along the index $x_v$ are given by 
\[
\Phi(v)_{x_v=0} = \Phi((v,\low(v))), \quad
\Phi(v)_{x_v=1} = \Phi((v,\high(v))).
\]
Using a slight extension of the Boolean–Shannon decomposition, we can write 
\begin{equation}\label{eq:limtdd-intp}
\Phi(v)
= \overline{x}_v \cdot \Phi((v,\low(v)))
+ x_v \cdot \Phi((v,\high(v))).
\end{equation}
}

For simplicity, for a rank-$n$ tensor $\phi$ with indices $x_n, \ldots, x_1$, we always adopt the ordering $x_n\prec \cdots \prec x_1$, which implies that $x_n$ is above $x_1$ in the LimTDD representation of $\phi$. 
\addt{
\begin{exam}
Fig.~\ref{fig:run_exp}\,(c) shows the LimTDD $\F$ of the tensor 
$\phi_{x_3x_2x_1}$ introduced in Example~\ref{exp:tensor}. For convenience, we write $v_i$ ($i=1,2,3$) for the node labeled with $x_i$ in $\F$ and write $\Phi(v_i)$ for the tensor represented by $v_i$. 

Recall that the tensor corresponds to the terminal node $v_T$ is 1. As  $\low(v_1)=v_T$ and the weight on the low edge of $v_1$ is 1, by Eq.~\ref{eq:vecc-edge}, we have 
\[\Phi((v_1,\low(v_1)))=\text{wt}((v_1,\low(v_1)))\cdot\Phi(v_T) = 1\cdot 1 = 1.\]
Similarly, $\Phi((v_1,\high(v_1)))=1$. We have by Eq.~\ref{eq:limtdd-intp}
\[ \Phi(v_1)=\overline{x_1}\cdot \Phi((v_1,\low(v_1))+x_1\cdot\Phi(v_1,\high(v_1)) =\overline{x_1} + x_1.\]
 This implies that $\Phi(v_1)(0)=1$ and $\Phi(v_1)(1)=1$. Thus tensor $\Phi(v_1)$ has vector representation $\vecc{\Phi(v_1)} = [1,\,1]^\intercal$.

For node $v_2$, we have $\Phi((v_2,\low(v_2))=\Phi(v_1)$, $\Phi((v_2,\high(v_2))=P^4\cdot\Phi(v_1)=Z\cdot \Phi(v_1)$, and 
\[\Phi(v_2)=\overline{x_2}\cdot\Phi(v_1)+x_2\cdot Z\cdot \Phi(v_1).\]
As $\vecc{\Phi(v_1)} = [1,\,1]^\intercal$ and  
$\vecc{Z\cdot \Phi(v_1)} = Z [1,\,1]^\intercal = [1,\,-1]^\intercal$, we have
$\vecc{\Phi(v_2)} = [1,\,1,\,1,\,-1]^\intercal$.

For node $v_3$, we have 
$\Phi((v_3,\low(v_3))=\Phi(v_2)$, $\Phi((v_3,\high(v_3))=(P^6\otimes P^4)\cdot\Phi(v_2)=(ZS\otimes Z)\cdot \Phi(v_2)$, and 
\[\Phi(v_3)=\overline{x_3}\cdot\Phi(v_2)+x_3\cdot (ZS\otimes Z)\cdot \Phi(v_2).\]
Because 
$\vecc{(ZS\otimes Z)\cdot \Phi(v_2)} = (ZS\otimes Z) [1,\,1,\,1,\,-1]^\intercal = [1,\,-1,\,-i,\,-i]^\intercal$, we have  
$\vecc{\Phi(v_3)} = [1,\,1,\,1,\,-1,\,1,\,-1,\,-i,\,-i]^\intercal$.

Finally, the tensor represented by $\mathcal{F}$ is the tensor $\Phi(\F)=\texttt{wt}(\F)\cdot\Phi(v_3)$, whose vector representation  is
\begin{align*}
\vecc{\Phi(\F)} = & \texttt{wt}(\F)\cdot \vecc{\Phi(v_3)} \\
=& \tfrac{1}{2\sqrt{2}}(P^6 \!\otimes\! I \!\otimes\! I)[1,\,1,\,1,\,-1,\,1,\,-1,\,-i,\,-i]^\intercal \\
=& \tfrac{1}{2\sqrt{2}}[1,\,1,\,1,\,-1,\,-i,\,i,\,-1,\,-1]^\intercal. 
\end{align*}
This is exactly the vector representation of $\phi_{x_3x_2x_1}$.
\end{exam}
}

Typically, a node $v$ in a LimTDD is uniquely determined by its index, its two successors, and the weights on the two edges connecting to those successors. Following \cite{vinkhuijzen2023limdd}, the LimTDD that represents the tensor of $v$ is denoted as 
\[\Circled{v_{0}}\overset{w_0}{\dashleftarrow}\Circled{v}\xrightarrow[]{w_1}\Circled{v_{1}}.\] 
Similarly, a LimTDD is uniquely determined by its root node and the weight on the incoming edge, represented as: 
\[\Big(w_\F,\ \Circled{v_{0}}\overset{w_0}{\dashleftarrow}\Circled{v}\xrightarrow[]{w_1}\Circled{v_{1}}\Big),\] or simply as \[\xrightarrow[]{w_\F}\Circled{v}.\] 



When the group $\mathcal{G}$ is restricted to the Pauli group, 
LimTDD and LIMDD become equivalent in representing quantum states. 
Furthermore, when $\mathcal{G}$ is restricted to scalar multiples of the identity 
(i.e., $\mathcal{G} \subset \mathbb{C} \cdot I$), 
LimTDD reduces exactly to the standard TDD. 
In our practical implementations, we choose $\mathcal{G}$ to be the $XP$-stabilizer group introduced in Sec.~\ref{sec:xplim}.

\begin{remark}
LIMDDs exploit the algebraic properties of Pauli operators to achieve compact representations and efficient computations. In LimTDDs, we extend these to $XP$-operators~\cite{webster_xp_2022}, enabling more compact decision diagrams while preserving canonicity through the algebraic structure of $\mathcal{XP}$. This extension also ensures efficient operations, such as slicing, which extracts sub-tensors from the diagram. 

We explain the rationale for selecting $\mathcal{XP}$ over more general groups for LimTDD. Several factors must be considered to ensure canonical representations and efficient computations in the decision diagram: 
\begin{enumerate}
    \item The operators must have a canonical form, with multiplication and inversion operations being computationally efficient.
    \item These operators should be efficiently extractable from the decision diagram structure. For example, an \( X \) operator can be extracted by swapping a node’s outgoing edges, while a \( P \) operator can be applied by adjusting the phase of the high edge.
    \item Computing the stabilizer group for a node (cf. Eq.~\ref{eq:stab_state}) within this group must be efficient to maintain the diagram’s canonical form; otherwise, adding stabilizers to edges without altering functionality could disrupt canonicity.
    \item The stabilizer group should have a compact representation.
    \item Computing sub-diagrams representing sub-tensors must be straightforward, as LimTDDs’ core algorithms, such as addition and contraction, rely on this operation.
\end{enumerate}
Fortunately, $XP$-operators satisfy all these requirements \cite{webster_xp_2022}.

\end{remark}


\subsection{LimTDD Normalization}


In decision diagrams with weighted paths, the function value associated with each path is obtained by multiplying the weights along that path. As a result, multiple decision diagrams with different weight assignments can represent the same function. \addt{The following lemma provides examples of LimTDDs, in general forms, that represent the same tensor.}



\begin{lem}
Suppose $v$ is a node in a LimTDD with form  
\[\Circled{v_{0}}\overset{w_0}{\dashleftarrow}\Circled{v}\xrightarrow[]{w_1}\Circled{v_{1}}.\] 
Let $g_i$ be a $XP$-stabilizer of $\vecc{v_i}$ $(i=0,1)$, and $0\leq k <N$.
Then the following LimTDDs represent the same tensor $\Phi(v)$:
\[\Big( 1 ,\  \Circled{v_{0}}\overset{w_0\cdot g_0}{\dashleftarrow}\Circled{v}\xrightarrow[]{w_1\cdot g_1}\Circled{v_{1}}\Big),\] 
\[\Big(P^ k \otimes (w_0\cdot g_0),\  \Circled{v_{0}}\overset{I}{\dashleftarrow}\Circled{v}\xrightarrow[]{ \omega^{-2k}\cdot g_0^\dagger\cdot w_0^\dagger\cdot w_1\cdot g_1}\Circled{v_{1}}\Big),\] 
\[\Big( X \otimes (w_1\cdot g_1),\  \Circled{v_{1}}\overset{I}{\dashleftarrow}\Circled{v}\xrightarrow[]{g_1^\dagger\cdot w_1^\dagger\cdot w_0\cdot g_0}\Circled{v_{0}}\Big).\]

\end{lem}

\addt{In the above result, the successors $v_0,v_1$ can represent arbitrary tensors. We next give a concrete example, where $v_0,v_1$ represent rank 1 tensors.}


\begin{exam}\label{exp:id_dds}
 \addt{Consider a node $\Big( 1 ,\  \Circled{v_{0}}\overset{I}{\dashleftarrow}\Circled{v}\xrightarrow[]{I}\Circled{v_{1}}\Big)$. Suppose $\vecc{v_0}=[1,0]^\intercal, \vecc{v_1} = [1,1]^\intercal$. The vectorization of $\Phi(v)$ is $[1,0,1,1]^\intercal$, where $\Phi(v)$ is the tensor corresponds to node $v$. Because $P^k[1,0]^\intercal=[1,0]^\intercal$ for any $k \in \mathbb{N}$ and $X[1,1]^\intercal=[1,1]^\intercal$, we know $\Big( 1 ,\  \Circled{v_{0}}\overset{P^k}{\dashleftarrow}\Circled{v}\xrightarrow[]{X}\Circled{v_{1}}\Big)$ represents $\Phi(v)$.
 As $X\otimes I [1,1,1,0]^\intercal = [1,0,1,1]^\intercal$, $\Big( X\otimes I ,\  \Circled{v_{1}}\overset{I}{\dashleftarrow}\Circled{v}\xrightarrow[]{I}\Circled{v_{0}}\Big)$ also
 represents $\Phi(v)$.} 

\end{exam}

The above lemma and example show that the same tensor can be represented by several different LimTDDs. Thus, a normalization process should be conducted. 

For any node $v$ in a LimTDD, we write

\begin{equation}\label{eq:stab_v}
\stab(v) = \{O\in \mathcal{XP}\ |\ O\vecc{\Phi(v)} =\vecc{\Phi(v)}\}  
\end{equation}
for the set of stabilizers of $v$ in $\mathcal{XP}$. The procedure for calculating $\stab(v)$ is described in \ref{sec:stab}. The complexity for calculating $\stab(v)$ is polynomial to the rank of the tensor represented by $v$.

\addt{Given a node $\Circled{v_{0}}\overset{w_0}{\dashleftarrow}\Circled{v}\xrightarrow[]{w_1}\Circled{v_{1}}\Big)$, to ensure a unique representation, two sets of weights need to be considered:} 
\begin{eqnarray*}\label{eq:W0}
    W_0 := \{g_0^{\dag} \cdot w_{0}^\dag \cdot w_{1} \cdot g_1 \ |\ g_0 \in \stab(v_0), g_1 \in \stab(v_1)\},  \\ \label{eq:W1}
    W_1 :=
    \{g_1^{\dag} \cdot w_{1}^\dag \cdot w_{0} \cdot g_0\ |\ g_0 \in \stab(v_0), g_1 \in \stab(v_1)\}. 
\end{eqnarray*}

For LimTDD, the normalization process incorporates three principles  to ensure a unique representation for each internal node $v$:
\begin{enumerate}
    \item $\low(v) \leq \high(v)$.
    \item $\w((v,\low(v))) =I$, unless it is 0. 
    \item If $\low(v) < \high(v)$, then $\w((v,\high(v)))$ is the smallest one in $W_1$, up to a complex coefficient; otherwise, $\low(v) = \high(v)$ and $\w((v,\high(v)))$ is the smallest in $W_0\cup W_1$, up to a complex coefficient. 
\end{enumerate}
Here, for two nodes $v_1, v_2$, $v_1 < v_2$ if $v_1$ is created before $v_2$ (typically with a smaller id). For two weights $w_1 = r_1 e^{2\pi i \theta_1} XP_N(p_1|\bold{x_1}|\bold{z_1})$ and $w_2 = r_2e^{2\pi i \theta_2}XP_N(p_2|\bold{x_2}|\bold{z_2})$,
\[w_1<w_2\quad \text{if}\quad (\bold{x_1}|\bold{z_1}|r_1|\theta_1|p_1) < (\bold{x_2}|\bold{z_2}|r_2|\theta_2|p_2)\] 
in lexicographical order, where $r_1, r_2, \theta_1, \theta_2 \in \mathbb{R}$, and $r_1,r_2\geq 0$. $r_1 e^{2\pi i \theta_1}$ and $r_2 e^{2\pi i \theta_2}$ are the polar forms of complex coefficients \addt{(cf. the notations after Definition~\ref{def:lim})}.

\begin{algorithm}
\caption{$\locnorm(x,\F_0,\F_1)$}
\begin{algorithmic}[1]
\Require{Two normalized LimTDDs $\xrightarrow[]{w_i}\Circled{v_i}$ ($i=0,1)$; and an index $x$. We require $v_0\leq v_1$}. 

\Ensure{A normalized LimTDD $\Big(\tilde{w},\Circled{v_{0}}\overset{I}{\dashleftarrow}\Circled{v}\xrightarrow[]{\hat{w}}\Circled{v_{1}}\Big)$}.
\State $b_n = 0$\hfill \# \texttt{Are $v_0,v_1$ exchanged?}
\If{$w_0=w_1=0$ }
\State \Return $\texttt{make\_dd}(0,x,0,v_0,0,v_1)$  
\EndIf
\State $w_\text{temp} \leftarrow \min_{g_0 \in \stab(v_0),g_1 \in \stab(v_1)} {g_0^{\dag} \cdot w_0^\dag \cdot w_1 \cdot g_1}$
\State $w_\text{temp}' \leftarrow \min_{g_0 \in \stab(v_0),g_1 \in \stab(v_1)} {g_1^{\dag} \cdot w_1^\dag \cdot w_0 \cdot g_0}$
\If{$v_1 = v_0$ and $w_\text{temp}' < w_\text{temp}$}
\State $b_n \leftarrow 1$
\State $g_1^\star \leftarrow \text{the stabilizer of } v_1 \text{ yielding } w_\text{temp}'$
\State $w \leftarrow w_1\cdot g_1^\star$ \hfill 
\State $w'_1 \leftarrow w_\text{temp}'$

\Else
\State $g_0^\star \leftarrow \text{the stabilizer of } v_0 \text{ yielding } w_\text{temp}$
\State $w \leftarrow w_0 \cdot g_0^\star$ \hfill 
\State $w'_1 \leftarrow w_\text{temp}$
\EndIf


\State $k \leftarrow \lfloor \texttt{ang}(w'_1)/2\omega \rfloor$
\State $dd \leftarrow \texttt{make\_dd}(X^{b_n}P^k \otimes w,x,I,v_0,w'_1/\omega^{2k},v_1)$
\State \Return $\texttt{check\_unique\_table}(dd)$
\end{algorithmic}
\label{alg:LimTDD_norm} 
\end{algorithm}

The normalization process adopts a bottom-up approach, enforcing the three conditions on every node through the $\locnorm$ procedure (cf. Algorithm~\ref{alg:LimTDD_norm}). This procedure takes two sub-LimTDDs, $\F_0 = \xrightarrow[]{w_0}\Circled{v_0}$ and $\F_1 = \xrightarrow[]{w_1}\Circled{v_1}$, and the index of their intended parent node $x$ as inputs. Assume $v_0\leq v_1$; that is, $v_0<v_1$ or they are identical. Otherwise, we swap the two nodes and add a $X$ operator. If $v_0<v_1$, i.e., $v_0$ is created before $v_1$, we set $\low(v)=v_0$ and $\high(v)=v_1$, and select the smallest value in $W_1$ as the weight on the high edge; otherwise, $v_0=v_1$ and we select the smallest value $w'_1$ in $W_0\cup W_1$ as the weight on the high edge. After that, the weight of the low-edge will be set to $I$ (unless it is 0), and the weight of the resulting LimTDD will be adjusted correspondingly. 


The subroutine $\texttt{make\_dd}(w,x,w_0,v_0,w_1,v_1)$ in this algorithm is used to construct a LimTDD $$\Big( w ,\  \Circled{v_{0}}\overset{w_0}{\dashleftarrow}\Circled{v}\xrightarrow[]{w_1}\Circled{v_{1}}\Big).$$
In the last step of the procedure, we check if the node of the resulted LimTDD already exists in the \texttt{unique\_table}; if so, reuse the node; otherwise, we generate a new LimTDD node and store it in the hash table with the key $(x,w_0,r_{\F_0},w_1,r_{\F_1})$. With these techniques, we make the LimTDD unique. 
\begin{exam}
 \addt{
 Suppose $\xrightarrow[]{P^k}\Circled{v_0}$ and $\xrightarrow[]{X}\Circled{v_1}$ are two LimTDDs, where $\vecc{v_0} = [1,0]^\intercal$ and $\vecc{v_1}=[1,1]^\intercal$. We combine them to form a larger tensor using the $\locnorm$ operation. The resulting node is $\Big( 1 ,\  \Circled{v_{0}}\overset{I}{\dashleftarrow}\Circled{v}\xrightarrow[]{I}\Circled{v_{1}}\Big)$ if $v_0\leq v_1$, or $\Big( X \otimes I,\  \Circled{v_{1}}\overset{I}{\dashleftarrow}\Circled{v}\xrightarrow[]{I}\Circled{v_{0}}\Big)$ if $v_0> v_1$. Both representations are equivalent to $\Big( 1 ,\  \Circled{v_{0}}\overset{P^k}{\dashleftarrow}\Circled{v}\xrightarrow[]{X}\Circled{v_{1}}\Big)$, describing the same tensor.
 
 }
    
\end{exam}

New local normalizations can also be obtained from established ones. 
\begin{lem}\label{lem:norm}
    Suppose $\locnorm\Big(x,\xrightarrow[]{w_0}\Circled{v_0},\xrightarrow[]{w_1}\Circled{v_1}\Big) = \xrightarrow[]{w}\Circled{v}$. Then we have the following new local normalizations:
    \begin{itemize}
    \item[(i)] $\locnorm\Big(x,\xrightarrow[]{O\cdot w_0}\Circled{v_0},\xrightarrow[]{O\cdot w_1}\Circled{v_1}\Big) = \xrightarrow[]{I\otimes O\cdot w}\Circled{v}$, for any $O \in \mathcal{XP}$; 
    \item[(ii)]  $\locnorm\Big(x,\xrightarrow[]{w_0}\Circled{v_0},\xrightarrow[]{\omega^{2k}w_1}\Circled{v_1}\Big) = \xrightarrow[]{(P^k\otimes I)\cdot w}\Circled{v}$;
    \item[(iii)]  $\locnorm\Big(x,\xrightarrow[]{w_1}\Circled{v_1},\xrightarrow[]{w_0}\Circled{v_0}\Big) =\xrightarrow[]{(X\otimes I) \cdot w}\Circled{v}$, if $v_0\neq v_1$ or $w_0 \neq w_1$;
    \item[(iv)]  $\locnorm\Big(x,\xrightarrow[]{w_0 \cdot g_0}\Circled{v_0},\xrightarrow[]{w_1 \cdot g_1}\Circled{v_1}\Big) = \xrightarrow[]{w}\Circled{v}$, for any $g_0 \in \stab(v_0)$ and $g_1 \in \stab(v_1)$.
    
    \end{itemize} 
\end{lem}

\addt{This lemma provides further opportunities for merging nodes. Suppose, for example, we are given two sub-LimTDDs, $\xrightarrow[]{O\cdot w_0}\Circled{v_0}$ and $\xrightarrow[]{O\cdot w_1}\Circled{v_1}$. The normalization result share the same node $v$ obtained from normalizing $\xrightarrow[]{w_0}\Circled{v_0}$ and $\xrightarrow[]{w_1}\Circled{v_1}$.
}

\begin{remark}
\addt{
The $\locnorm$ operation corresponds to the $MakeEdge$ operation described in~\cite{vinkhuijzen2023limdd}. However, computing $\stab(v)$ for $XP$-operators, as detailed in Appendix~\ref{sec:stab}, is more complex than for XZ-stabilizers. Additionally, for any $k \in \mathbb{N}$, we can extract the phase $\omega^{2k}$ from the weight of a node’s high edge and replace it with a $P^k$ operator on the incoming edge, whereas only the phase $-1$ can be extracted for XZ-stabilizers. This provides more opportunity for merging two nodes (cf. Lemma~\ref{lem:norm} (ii)). 
}
\end{remark}

\section{Algorithms}\label{sec:Algorithms}

This section presents algorithms for constructing LimTDDs from tensors and performing key operations such as addition and contraction. All algorithms are implemented recursively. As with TDDs, we employ local normalization whenever a node is created. Additionally, we use \texttt{computed\_table} (a hash table technique) to prevent redundant calculations. 

\addt{These algorithms have standard counterparts in traditional decision diagrams: slicing corresponds to restriction, addition corresponds to apply, and contraction corresponds to matrix multiplication (cf. \cite[Sec.~3.2]{bahar1997algebric}). However, unlike matrix multiplication, contraction allows us to handle operations on data with incompatible dimensions more easily. Here, we provide a concise restatement of these algorithms from a tensor network perspective. This approach introduces operator actions on edges, while also allowing users familiar with tensor networks to recognize and leverage these operations effectively.}

\subsection{LimTDD Generation}

To generate the LimTDD representation of a tensor, we recursively construct the LimTDD representations of its two sub-tensors and then apply the normalization procedure. For a scalar, the representation is a LimTDD with a single terminal node weighted by that scalar. This process aligns with the TDD generation method described in~\cite{hong2020tensor} (see Algorithm \ref{alg:LimTDD_generate}). The construction has a time complexity linear in the tensor's size (i.e., the number of values in the tensor).

\begin{algorithm}
\caption{$\LimTDD\_generate(\phi)$}
\begin{algorithmic}[1]
\Require{A tensor $\phi$ over a linearly ordered index set $S$.}
\Ensure{The normalized LimTDD of $\phi$.}

\If{$\phi \equiv c$ is a constant}
\State \Return $\xrightarrow[]{\ c\ }\Circled{v_T}$
\EndIf
\State  $x\leftarrow$ the smallest index of $\phi$
\State $r_0 \leftarrow \LimTDD\_generate(\phi_{x=0})$
\State $r_1 \leftarrow \LimTDD\_generate(\phi_{x=1})$
\State \Return $\locnorm(x,r_0,r_1)$
\end{algorithmic}
\label{alg:LimTDD_generate}
\end{algorithm}


\subsection{LimTDD Slicing}


\addt{LimTDD operations, such as addition and contraction, rely on a slicing step, which involves extracting the sub-LimTDD of a sub-tensor from a LimTDD.} Let $\mathcal{F}$ be the LimTDD representation of a tensor $\phi_{x_n \cdots x_1}$. In this section, we describe how to extract from $\mathcal{F}$ the LimTDDs $\mathcal{F}{x_i = c}$ that represent the tensors $\phi_{x_i=c}$ for $c \in \{0,1\}$. Unlike TDDs, this process must account for edge operators, ensuring their effects are propagated to sub-LimTDDs.


First,  consider the case when $x_i=x_n$ is the top index. Suppose $\F = \Big(w_\F,\Circled{v_{0}}\overset{w_0}{\dashleftarrow}\Circled{v}\xrightarrow[]{w_1}\Circled{v_{1}}\Big)$, with $w_\F= X^{b_n}P^{z_n}\otimes w$. Here, $P^{z_n}$ adds a phase $\omega^{2z_n}$ to the high edge's weight, and $X^{b_n}$ swaps the 0-successor and 1-successor if $b_n=1$. Thus, 
\begin{itemize}
    \item if $b_n=0$, then $\F_{x_i = c} = \xrightarrow[]{w \cdot w_{c}}\Circled{v_{c}}$, and 
    \item if $b_n=1$, then $\F_{x_i = c} = \xrightarrow[]{w \cdot w_{1-c}}\Circled{v_{1-c}}$. 
\end{itemize}
Note that $w_1 = \omega^{2z_n}w_1$ in the high edge's weight. 

If $x_i\neq x_n$ (i.e., $x_i$ is not the top index), we first calculate $\F_{x_n = b}$ for $b \in \{0,1\}$, then recursively calculate $(\F_{x_n = b})_{x_i=c}$, and combine the results. If $x_i$ is not an index of $\F$, $\F_{x_n = c}=\F$ for both $c=0$ and $1$.

\begin{algorithm}
\caption{$\texttt{Slicing}(\F,x,c)$}
\begin{algorithmic}[1]
\Require{A LimTDD $\F$ with weight $w_\F = \lambda X^{b_n}P^{z_n}\otimes w$ representing the tensor $\phi$; an index $x$ and a value $c\in\set{0,1}$.}
\Ensure{The sub-LimTDD of $\F$ that represents $\phi_{x=c}$.}

\If{$\F$ is a trivial LimTDD}
\State \Return $\F$
\EndIf

\State $x^{\prime}\leftarrow \idx(r_{\F})$

\If{$x < x^{\prime}$}
\State \Return $\F$
\EndIf

\If{$x = x^{\prime}$}
\State $dd\leftarrow$ an empty LimTDD
\State $v_0 \leftarrow \low(r_\F)$
\State $v_1 \leftarrow \high(r_\F)$
\State $w_0 \leftarrow \w(r_\F,\low(r_\F))$
\State $w_1 \leftarrow \w(r_\F,\high(r_\F))$
\If{$c=0$}
\State $dd.node = v_{b_n}$
\State $dd.weight = \lambda \cdot \omega^{2z b_n} \cdot w \cdot  w_{b_n}$
\Else
\State $dd.node = v_{1-b_n}$
\State $dd.weight = \lambda \cdot \omega^{2z (1-b_n)} \cdot  w \cdot w_{1-b_n}$
\EndIf
\State \Return $dd$
\EndIf

\If{$x > x^{\prime}$}
\State $r_0 \leftarrow \texttt{Slicing}(\texttt{Slicing}(\F,x^{\prime},0),x,c)$
\State $r_1 \leftarrow \texttt{Slicing}(\texttt{Slicing}(\F,x^{\prime},1),x,c)$
\State \Return $\locnorm(x^{\prime},r_0,r_1)$
\EndIf

\end{algorithmic}
\label{alg:LimTDD_Slicing}
\end{algorithm}


The general slicing procedure for LimTDD is similar to that for TDD described in \cite{hong2020tensor} (see Algorithm \ref{alg:LimTDD_Slicing}). This procedure extracts the portion of data in $\Phi(\F)$ with $x_i=c$ and leads to the following result:

\begin{lem}\label{lem:sub-limtdd}
    Let $\F$ be the LimTDD of tensor $\phi_{x_n\cdots x_1}$. For $c\in \{0,1\}$, define $\F_{x_i=c} = \texttt{Slicing}(\F,x_i,c)$. Then, $\F_{x_i=c}$ is the LimTDD representation of the tensor $\phi_{x_i=c}$.
\end{lem}

For the root node of a LimTDD, the \texttt{Slicing} operation is the inverse operation of $\locnorm$: applying the $\locnorm$ operation to the two sliced sub-LimTDDs with respect to $x_n$ retrieves $\F$.

\begin{lem}\label{lem:sub-limtdd_com}
Let $\F$ be the LimTDD representation of tensor $\phi_{x_n\cdots x_1}$. Then 
    $\locnorm(x_n,\F_{x_n=0},\F_{x_n=1}) = \F$.
\end{lem}

\addt{With the slicing operation defined, we now introduce the core LimTDD operations, namely addition and contraction. These operations enable LimTDD to address most tensor network-based tasks.}

\subsection{LimTDD Addition}

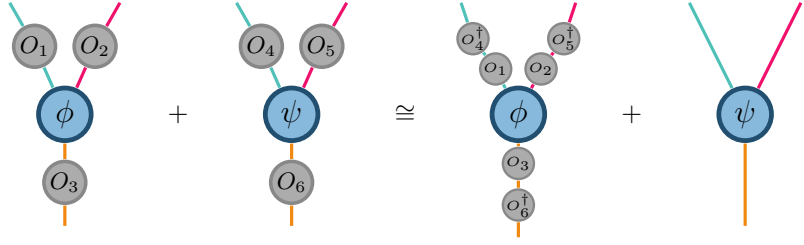
\begin{figure*}[!tbh]
    \centering
\usetikzlibrary{backgrounds,fit,calc}
\definecolor{tensorBlue}{HTML}{224f70}
\definecolor{tensorBlueLight}{HTML}{87B9DD}
\definecolor{nodeGray}{HTML}{888888}
\definecolor{nodeGrayLight}{HTML}{ACACAC}
\definecolor{edgeOrange}{HTML}{f18810}
\definecolor{edgeMagenta}{HTML}{f0116d}
\definecolor{edgeGreen}{HTML}{50c1b8}


\begin{tikzpicture}[
  tensor ring/.style args={#1/#2}{
    circle, fill=#1, minimum size=4.8ex, inner sep=0pt,
    append after command={
      \pgfextra{\let\mytikzlastnode\tikzlastnode}
      node[circle, fill=#2, minimum size=4ex, inner sep=0pt] at (\mytikzlastnode) {}
    }
  },
  small gray ring/.style={
    circle, fill=nodeGray, minimum size=3.8ex, inner sep=0pt,
    append after command={
      \pgfextra{\let\mytikzlastnode\tikzlastnode}
      node[circle, fill=nodeGrayLight, minimum size=3.2ex, inner sep=0pt] at (\mytikzlastnode) {}
    }
  },
  tiny gray ring/.style={
    circle, fill=nodeGray, minimum size=2.8ex, inner sep=0pt,
    append after command={
      \pgfextra{\let\mytikzlastnode\tikzlastnode}
      node[circle, fill=nodeGrayLight, minimum size=2.4ex, inner sep=0pt] at (\mytikzlastnode) {}
    }
  },
  tensor text/.style={inner sep=0pt, font=\large},
  small text/.style={inner sep=0pt, font=\small},
  tiny text/.style={inner sep=0pt, font=\tiny},
]
  \begin{scope}
    \node[tensor ring=tensorBlue/tensorBlueLight] (phi) at (0,0) {};
    \node[small gray ring] (O1) at (-0.4,0.9) {};
    \node[] (O1End) at (-0.8,1.6) {};
    \node[small gray ring] (O2) at (0.4,0.9) {};
    \node[] (O2End) at (0.8,1.6) {};
    \node[small gray ring] (O3) at (0,-0.9) {};
    \node[] (O3End) at (0,-1.6) {};
    
    \node[tensor text] at (phi) {$\phi$};
    \node[small text] at (O1) {$O_1$};
    \node[small text] at (O2) {$O_2$};
    \node[small text] at (O3) {$O_3$};
    
    \draw[very thick,edgeGreen] (phi) -- (O1);
    \draw[very thick,edgeMagenta] (phi) -- (O2);
    \draw[very thick,edgeOrange] (phi) -- (O3);
    \draw[very thick,edgeGreen] (O1End) -- (O1);
    \draw[very thick,edgeMagenta] (O2End) -- (O2);
    \draw[very thick,edgeOrange] (O3End) -- (O3);
  \end{scope}

  \begin{scope}[shift={(1.5,0)}]
    \node at (0,0) {$+$};
  \end{scope}

  \begin{scope}[shift={(3,0)}]
    \node[tensor ring=tensorBlue/tensorBlueLight] (psi) at (0,0) {};
    \node[small gray ring] (O4) at (-0.4,0.9) {};
    \node[] (O4End) at (-0.8,1.6) {};
    \node[small gray ring] (O5) at (0.4,0.9) {};
    \node[] (O5End) at (0.8,1.6) {};
    \node[small gray ring] (O6) at (0,-0.9) {};
    \node[] (O6End) at (0,-1.6) {};
    
    \node[tensor text] at (psi) {$\psi$};
    \node[small text] at (O4) {$O_4$};
    \node[small text] at (O5) {$O_5$};
    \node[small text] at (O6) {$O_6$};
    
    \draw[very thick,edgeGreen] (psi) -- (O4);
    \draw[very thick,edgeMagenta] (psi) -- (O5);
    \draw[very thick,edgeOrange] (psi) -- (O6);
    \draw[very thick,edgeGreen] (O4End) -- (O4);
    \draw[very thick,edgeMagenta] (O5End) -- (O5);
    \draw[very thick,edgeOrange] (O6End) -- (O6);
  \end{scope}

  \begin{scope}[shift={(4.5,0)}]
    \node at (0,0) {$\cong $};
  \end{scope}
  
  \begin{scope}[shift={(6,0)}]
    \node[tensor ring=tensorBlue/tensorBlueLight] (phi) at (0,0) {};
    \node[tiny gray ring] (O1) at (-0.3,0.6) {};
    \node[tiny gray ring] (O4dagger) at (-0.6,1) {};
    \node[] (O4daggerEnd) at (-0.8,1.6) {};

    \node[tiny gray ring] (O2) at (0.3,0.6) {};
    \node[tiny gray ring] (O5dagger) at (0.6,1) {};
    \node[] (O5daggerEnd) at (0.8,1.6) {};

    \node[tiny gray ring] (O3) at (0,-0.65) {};
    \node[tiny gray ring] (O6dagger) at (0,-1.2) {};
    \node[] (O6daggerEnd) at (0,-1.75) {};
    
    \node[tensor text] at (phi) {$\phi$};
    \node[tiny text] at (O1) {$O_1$};
    \node[tiny text] at (O4dagger) {$O_4^\dagger$};
    \node[tiny text] at (O2) {$O_2$};
    \node[tiny text] at (O5dagger) {$O_5^\dagger$};
    \node[tiny text] at (O3) {$O_3$};
    \node[tiny text] at (O6dagger) {$O_6^\dagger$};

    \draw[very thick,edgeGreen] (phi) -- (O1);
    \draw[very thick,edgeMagenta] (phi) -- (O2);
    \draw[very thick,edgeOrange] (phi) -- (O3);

    \draw[very thick,edgeGreen] (O1) -- (O4dagger);
    \draw[very thick,edgeGreen] (O4daggerEnd) -- (O4dagger);

    \draw[very thick,edgeMagenta] (O2) -- (O5dagger);
    \draw[very thick,edgeMagenta] (O5daggerEnd) -- (O5dagger);

    \draw[very thick,edgeOrange] (O3) -- (O6dagger);
    \draw[very thick,edgeOrange] (O6daggerEnd) -- (O6dagger);
  \end{scope}

  \begin{scope}[shift={(7.5,0)}]
    \node at (0,0) {$+ $};
  \end{scope}

  \begin{scope}[shift={(9,0)}]
    \node[tensor ring=tensorBlue/tensorBlueLight] (psi) at (0,0) {};
    \node[] (O4End) at (-0.8,1.6) {};
    \node[] (O5End) at (0.8,1.6) {};
    \node[] (O6End) at (0,-1.6) {};
    
    \node[tensor text] at (psi) {$\psi$};
    
    \draw[very thick,edgeGreen] (psi) -- (O4End);
    \draw[very thick,edgeMagenta] (psi) -- (O5End);
    \draw[very thick,edgeOrange] (psi) -- (O6End);
  \end{scope}

\end{tikzpicture}

\caption{The addition of two tensors using LimTDD.}
\label{fig:limtdd_add}
\end{figure*}

Let $\F$ and $\G$ be two LimTDDs over an index set $S$. The sum of $\F$ and $\G$, denoted $\F+\G$, is a LimTDD representing the tensor $\bphi(\F)+\bphi(\G)$. Similar to TDD addition, the addition of two LimTDDs can be calculated recursively by adding their sub-LimTDDs. For any $x\in S$ where $x\preceq \idx(r_\F)$ and $x\preceq \idx(r_\G)$, the LimTDD version of the Boole-Shannon expansion gives us 
\begin{align*}
\bphi(\F)+ \bphi(\G) = &\  \overline{x} \cdot (\bphi(\F_{x=0})+ \bphi(\G_{x=0}))\\ 
&\quad\quad + x \cdot (\bphi(\F_{x=1}) + \bphi(\G_{x=1})).
\end{align*}
Here $\F_{x=c}$ and $\G_{x=c}$ are the sub-LimTDDs as defined in Lemma~\ref{lem:sub-limtdd} for $c\in \{0,1\}$.

\begin{algorithm}
\caption{$\add(\F, \G)$}
\begin{algorithmic}[1]
\Require{
Two LimTDDs $\F$ and $\G$}.
\Ensure {The LimTDD representation of $\bphi(\F)+\bphi(\G)$}.
\If{$r_\F= r_\G$ and $\w(\F) = c \cdot \w(\G)$}
\State $dd\leftarrow \F$
\State $dd.weight\leftarrow w_\F\ + \ w_\G$
\State \Return $dd$
\EndIf

\State $l \leftarrow$ the smaller one of $\F$ and $\G$
\State $h \leftarrow$ the bigger one of $\F$ and $\G$
\State $temp\_w \leftarrow \w(h)$
\State $\w(l) \leftarrow \w(h)^\dag\cdot \w(l)$
\State $\w(h) \leftarrow I$

\State $dd\leftarrow \texttt{find\_computed\_table}(l,h,+)$
\If {$dd$ is empty}
\State $x\leftarrow$ the smaller index of $r_\F$ and $r_\G$

\State $r_0 \leftarrow \add(l_{x=0},h_{x=0})$
\State $r_1 \leftarrow \add(l_{x=1},h_{x=1})$

\State $dd\leftarrow \locnorm(x,r_0,r_1)$
\EndIf
\State $dd.weight\leftarrow temp\_w \cdot dd.weight$
\State \Return $dd$

\end{algorithmic}
\label{alg:LimTDD_Add}
\end{algorithm}

The addition procedure for LimTDD is similar to that for TDD \cite{hong2020tensor} (see Algorithm \ref{alg:LimTDD_Add}).
To enhance the efficiency of reusing intermediate results, we first transfer the weights of one LimTDD to the other before performing the recursive step. As illustrated in Fig.~\ref{fig:limtdd_add}, to calculate the sum $O_3\otimes O_2\otimes O_1 \vecc{\phi}+O_{6}\otimes O_5 \otimes O_4\vecc{\psi}$, we first compute $(O_{6}\otimes O_5 \otimes O_4)^\dag \cdot O_3\otimes O_2 \otimes O_1 \vecc{\phi}+\vecc{\psi}$ and then multiply the result's weight by $O_{6}\otimes O_5 \otimes O_4$. Second, this approach enables reuse of the intermediate result when computing two tensors that differ only by a multiplicative factor. For example, when calculating $O\cdot O_3\otimes O_2 \otimes O_1 \vecc{\phi}+ O\cdot O_{6}\otimes O_5 \otimes O_4\vecc{\psi}$, the calculated sum $O_3\otimes O_2\otimes O_1 \vecc{\phi}+O_{6}\otimes O_5 \otimes O_4\vecc{\psi}$ can be reused, improving computational efficiency.

\addt{LimTDD performs well in practice, but, like TDD, its worst-case complexity depends on the number of non-zero paths in the two diagrams involved in the addition. Specifically, if $\mathcal{P}(\F)$ and $\mathcal{P}(\G)$ denote the number of non-zero paths in $\F$ and $\G$, the worst-case complexity is $\mathcal{O}(\mathcal{P}(\F) \cdot \mathcal{P}(\G))$. This contrasts with standard BDDs, where complexity typically depends only on the number of nodes. The reason is that, when adding two nodes $v_0$ and $v_1$ with weights $w_0$ and $w_1$, the operation cannot be reduced to simply computing $v_0 + v_1$ and then adding $w_0 + w_1$ to the edge weights.}



\subsection{LimTDD Contraction}




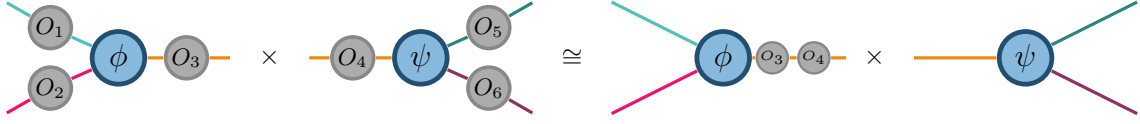
\begin{figure*}[tbh]
    \centering
\usetikzlibrary{backgrounds,fit,calc}
\definecolor{tensorBlue}{HTML}{224f70}
\definecolor{tensorBlueLight}{HTML}{87B9DD}
\definecolor{nodeGray}{HTML}{888888}
\definecolor{nodeGrayLight}{HTML}{ACACAC}
\definecolor{edgeOrange}{HTML}{f18810}
\definecolor{edgeMagenta}{HTML}{f0116d}
\definecolor{edgeGreen}{HTML}{50c1b8}
\definecolor{edgeDarkGreen}{HTML}{2c847d}
\definecolor{edgeDarkMagenta}{HTML}{8d315f}


\begin{tikzpicture}[
  tensor ring/.style args={#1/#2}{
    circle, fill=#1, minimum size=4.8ex, inner sep=0pt,
    append after command={
      \pgfextra{\let\mytikzlastnode\tikzlastnode}
      node[circle, fill=#2, minimum size=4ex, inner sep=0pt] at (\mytikzlastnode) {}
    }
  },
  small gray ring/.style={
    circle, fill=nodeGray, minimum size=3.8ex, inner sep=0pt,
    append after command={
      \pgfextra{\let\mytikzlastnode\tikzlastnode}
      node[circle, fill=nodeGrayLight, minimum size=3.2ex, inner sep=0pt] at (\mytikzlastnode) {}
    }
  },
  tiny gray ring/.style={
    circle, fill=nodeGray, minimum size=2.8ex, inner sep=0pt,
    append after command={
      \pgfextra{\let\mytikzlastnode\tikzlastnode}
      node[circle, fill=nodeGrayLight, minimum size=2.4ex, inner sep=0pt] at (\mytikzlastnode) {}
    }
  },
  tensor text/.style={inner sep=0pt, font=\large},
  small text/.style={inner sep=0pt, font=\small},
  tiny text/.style={inner sep=0pt, font=\tiny},
]
  \begin{scope}
    \node[tensor ring=tensorBlue/tensorBlueLight] (phi) at (0,0) {};
    \node[small gray ring] (O1) at (-0.9,0.4) {};
    \node[] (O1End) at (-1.6,0.8) {};
    \node[small gray ring] (O2) at (-0.9,-0.4) {};
    \node[] (O2End) at (-1.6,-0.8) {};
    \node[small gray ring] (O3) at (0.9,0) {};
    \node[] (O3End) at (1.6,0) {};
    
    \node[tensor text] at (phi) {$\phi$};
    \node[small text] at (O1) {$O_1$};
    \node[small text] at (O2) {$O_2$};
    \node[small text] at (O3) {$O_3$};
    
    \draw[very thick,edgeGreen] (phi) -- (O1);
    \draw[very thick,edgeMagenta] (phi) -- (O2);
    \draw[very thick,edgeOrange] (phi) -- (O3);
    \draw[very thick,edgeGreen] (O1End) -- (O1);
    \draw[very thick,edgeMagenta] (O2End) -- (O2);
    \draw[very thick,edgeOrange] (O3End) -- (O3);
  \end{scope}

  \begin{scope}[shift={(2,0)}]
    \node at (0,0) {$\times$};
  \end{scope}

  \begin{scope}[shift={(4,0)}]
    \node[tensor ring=tensorBlue/tensorBlueLight] (psi) at (0,0) {};
    \node[small gray ring] (O4) at (-0.9,0) {};
    \node[] (O4End) at (-1.6,0) {};
    \node[small gray ring] (O5) at (0.9,0.4) {};
    \node[] (O5End) at (1.6,0.8) {};
    \node[small gray ring] (O6) at (0.9,-0.4) {};
    \node[] (O6End) at (1.6,-0.8) {};
    
    \node[tensor text] at (psi) {$\psi$};
    \node[small text] at (O4) {$O_4$};
    \node[small text] at (O5) {$O_5$};
    \node[small text] at (O6) {$O_6$};
    
    \draw[very thick,edgeOrange] (psi) -- (O4);
    \draw[very thick,edgeDarkGreen] (psi) -- (O5);
    \draw[very thick,edgeDarkMagenta] (psi) -- (O6);
    \draw[very thick,edgeOrange] (O4End) -- (O4);
    \draw[very thick,edgeDarkGreen] (O5End) -- (O5);
    \draw[very thick,edgeDarkMagenta] (O6End) -- (O6);
  \end{scope}

  \begin{scope}[shift={(6,0)}]
    \node at (0,0) {$\cong $};
  \end{scope}
  
  \begin{scope}[shift={(8,0)}]
    \node[tensor ring=tensorBlue/tensorBlueLight] (phi) at (0,0) {};
    \node[] (O1End) at (-1.6,0.8) {};
    \node[] (O2End) at (-1.6,-0.8) {};
    \node[tiny gray ring] (O3) at (0.65,0) {};
    \node[tiny gray ring] (O4) at (1.2,0) {};
    \node[] (O4End) at (1.75,0) {};
    
    \node[tensor text] at (phi) {$\phi$};

    \node[tiny text] at (O3) {$O_3$};
    \node[tiny text] at (O4) {$O_4$};
    
    \draw[very thick,edgeOrange] (phi) -- (O3);
    \draw[very thick,edgeGreen] (O1End) -- (phi);
    \draw[very thick,edgeMagenta] (O2End) -- (phi);
    \draw[very thick,edgeOrange] (O3) -- (O4);
    \draw[very thick,edgeOrange] (O4End) -- (O4);
  \end{scope}

  \begin{scope}[shift={(10,0)}]
    \node at (0,0) {$\times$};
  \end{scope}

  \begin{scope}[shift={(12,0)}]
    \node[tensor ring=tensorBlue/tensorBlueLight] (psi) at (0,0) {};
    \node[] (O4End) at (-1.6,0) {};
    \node[] (O5End) at (1.6,0.8) {};
    \node[] (O6End) at (1.6,-0.8) {};
    
    \node[tensor text] at (psi) {$\psi$};
    
    \draw[very thick,edgeOrange] (psi) -- (O4End);
    \draw[very thick,edgeDarkGreen] (psi) -- (O5End);
    \draw[very thick,edgeDarkMagenta] (psi) -- (O6End);
  \end{scope}

\end{tikzpicture}

\caption{The contraction of two tensors using LimTDD, where $\times$ denotes tensor contraction.}
\label{fig:limtdd_cont}
\end{figure*}

Contraction is a fundamental operation in tensor networks. This subsection details how to efficiently implement contraction using LimTDDs. The procedure is also similar to that presented in \cite{hong2020tensor} for TDD.

Let $\F$ and $\G$ be two LimTDDs, $S$ the set of indices appearing in $\F$ or $\G$, and $var$ a subset of $S$ representing the indices to be contacted. We use  $\cont(\cdot)$ for both tensor and LimTDD contractions. For any $x\in S$ with $x\preceq \idx(r_\F)$ and $x\preceq \idx(r_\G)$, Eq.~\ref{eq:contdef} implies that if $x\in var$, then $\cont(\left(\bphi(\F),\bphi(\G), var\right)$ equals
	$$\sum_{c=0}^{1} \cont(\bphi(\F_{x=c}),\bphi(\G_{x=c}), var\backslash\{x\});$$
otherwise, it equals
\begin{align*}
\hfill \overline{x} \cdot \cont(\bphi(\F_{x=0}),\bphi(\G_{x=0}), var) \quad\quad\quad\quad \\
	   + \ x \cdot \cont(\bphi(\F_{x=1}),\bphi(\G_{x=1}), var).
\end{align*}
This means that calculating the contraction of $\F$ and $\G$ involves calculating the contraction of their sub-LimTDDs and then adding or combining the results.

\begin{algorithm}
\caption{$\cont(\F, \G, var)$}
\begin{algorithmic}[1]
\Require {Two LimTDDs $\F$ and $\G$; the set $var$ of indices to be contracted.}
\Ensure {The contracted LimTDD over $var$.}
\If{both $\F$ and $\G$ are trivial}
\State $dd\leftarrow \F$
\State $dd.weight \leftarrow w_\F \cdot w_\G \cdot 2^{len(var)}$
\State \Return $dd$
\EndIf
\State $temp\_w_0 =$ operators in $\w(\F)$ applied to open indices
\State $temp\_w_1 =$ operators in $\w(\G)$ applied to open indices
\State $\w(\G) = w(\G)/temp\_w_1$
\State $\w(\F) = w(\G)^\dag\cdot \w(\F)/temp\_w_0$
\State $\w(\G) = I$

\State $dd\leftarrow \texttt{find\_computed\_table}(\F,\G,\times )$
\If {$dd$ is empty}
\If{$\G$ is identity}
\State $dd\leftarrow \F$
\Else
\State $x\leftarrow$ the smaller index of $r_\F$ and $r_\G$
\If{$x\in var$}
\State $r_0 \leftarrow \cont(\F_{x=0},\G_{x=0},var\backslash\{x\})$
\State $r_1 \leftarrow \cont(\F_{x=1},\G_{x=1},var\backslash\{x\})$
\State  $dd\leftarrow \add(r_0,r_1)$
\Else
\State $r_0 \leftarrow \cont(\F_{x=0},\G_{x=0},var)$
\State $r_1 \leftarrow \cont(\F_{x=1},\G_{x=1},var)$
\State $dd \leftarrow \locnorm(x,r_0,r_1)$
\EndIf
\EndIf
\EndIf
\State $dd.weight = temp\_w_0\cdot temp\_w_1\cdot dd.weight$
\State \Return $dd$
\end{algorithmic}
\label{alg:LimTDD_Contraction}
\end{algorithm}

\addt{ The tensor network representation of operators in LimTDD improves efficiency by enabling the reuse of intermediate results, thereby reducing the computational complexity of LimTDD operations. For example, consider the contraction of two LimTDDs representing $(O_3 \otimes O_2 \otimes O_1) \vecc{\phi}$ and $(O_6 \otimes O_5 \otimes O_4) \vecc{\psi}$. As illustrated in Fig.~\ref{fig:limtdd_cont}, we first remove operators applied to open indices ($O_1, O_2; O_5, O_6$) and then transfer operators applied to contracted indices ($O_3; O_4$) to a single LimTDD. This strategy allows intermediate results to be reused when computing contractions of LimTDDs that differ only by operators on open indices (e.g., replacing $O_1$ with $O \cdot O_1$) or by operators on contracted indices where the differences cancel out (e.g., replacing $O_3$ and $O_4$ with $O_3 \cdot O^\dagger$ and $O \cdot O_4$, respectively).}  

Algorithm~\ref{alg:LimTDD_Contraction} provides the detailed procedure for LimTDD contraction. While LimTDD is generally more compact and efficient, like TDD, its worst-case time complexity for contraction depends on the number of non-zero paths in the diagrams involved. Let $\mathcal{P}(\F)$ and $\mathcal{P}(\G)$ denote the number of non-zero paths in $\F$ and $\G$, respectively. Then the worst-case complexity is $\mathcal{O}(\mathcal{P}(\F)^2 \cdot \mathcal{P}(\G)^2)$.

\section{Compactness}\label{sec:compare}



\blue{To analyze the compactness of LimTDD, we first characterize isomorphic tensors in terms of their LimTDD representations.}

Recall that $\mathcal{XP_N}$ denote the group of $XP$-stabilizers with precision $N$, where $\omega=e^{\pi i/N}$. We use $N$-LimTDD as a shorthand for $\mathcal{XP_N}$-LimTDD. Suppose $\F$ and $\G$ are two  $N$-LimTDDs on the same index set and with the same index order. We write $\F \cong \G$ if $r_\F =r_\G$. 
\begin{thm}\label{thm:iso_nodes}

Suppose $\phi_{x_n\cdots x_1}$ and $\gamma_{x_n\cdots x_1}$ are two tensors with the same index set. Let $\F$ and $\G$ be the $N$-LimTDD representations of $\phi_{x_n\cdots x_1}$ and $\gamma_{x_n\cdots x_1}$ with the same index order. The following two conditions are equivalent:
\begin{enumerate}
    \item $\phi_{x_n\cdots x_1}$ and $\gamma_{x_n\cdots x_1}$ are $\mathcal{XP_N}$-isomorphic.
\item $r_\F =r_\G$ and $w_\F = O \cdot w_\G \cdot g$ for some $O\in \mathcal{XP_N}$ and some  $g \in \stab(r_\G)$.
\end{enumerate} 
\end{thm}

This theorem establishes that two $\mathcal{XP_N}$-isomorphic tensors have isomorphic $N$-LimTDD representations. Consequently, the  canonicity of LimTDDs follows from this theorem.

As the number of operators grows with increasing $N$, more isomorphism operators can be extracted between two tensors. This leads to the following result:
\begin{cor}\label{cor:size}
For any tensor $\phi$ and any $N\geq 0$, let $\F$ be its $(N+1)$-LimTDD representation and $\G$ its $N$-LimTDD representation. Then, $size(\F)\leq size(\G)$, where $size(\F)$ and $size(\G)$ denote the number of nodes in $\F$ and $\G$, respectively.
\end{cor}


For quantum states, LIMDDs are equivalent to $2$-LimTDDs. For general tensors, $1$-LimTDDs provide a more compact representation than TDDs by enabling further compression of tensors differing only by a series of $X$ operators. With a slight abuse of notation, we denote TDDs as $0$-LimTDDs. \blue{The following result asserts that LimTDD is at least as compact as TDD (LIMDD) when representing tensors (quantum states).}
\begin{thm}[LimTDD vs. TDD \& LIMDD]
\begin{enumerate}
    \item Let $\F$ be the LimTDD representation and $\G$ the TDD representation of the same tensor $\phi_{x_n\cdots x_1}$ with the same index order. Then, $size(\F)\leq size(\G)$.
    \item Let $\F$ be the N-LimTDD ($N\geq 2$) representation and $\G$ the LIMDD representation of the same quantum state $\ket{\phi_{x_n\cdots x_1}}$ with the same index (qubit) order. Then, $size(\F)\leq size(\G)$.
\end{enumerate}
\end{thm}

\blue{We demonstrate below that the size gap can be exponential in certain cases.} 

\subsection{Tower LimTDD}\label{sec:tower}
A LimTDD is in \emph{tower form} (illustrated in Fig.~\ref{fig:run_exp}) if for every internal $v$, the 0- and 1-successors are identical, i.e., $\low(v)=\high(v)$. In such cases, the tensors represented by the low and high edges of any internal node $v$ differ only by a local operator. Conversely, if two tensors differ only by a local operator, they could be represented as two LimTDDs sharing the same root node. This leads to the following corollary:

\begin{cor}[Tower LimTDD]
A tensor $\phi_{x_n\cdots x_{1}}$ can be represented as a tower LimTDD if and only if for every $i \in \{ n, \cdots, 1\}$, there exists an operator  $O_i \in \mathcal{O}(i-1)$ such that:  $\vecc{\phi_{x_n\cdots x_{i}=0\cdots0 0}} = O_i \vecc{\phi_{x_n\cdots x_{i}=0\cdots0 1}}$.
\end{cor}

In \cite{vinkhuijzen2023limdd}, it is shown that all stabilizer states admit tower-form representations using LIMDDs. Below, we identify additional classes of quantum states that can be represented as tower-form LimTDDs: 


\begin{enumerate}
    \item States prepared from $\ket{0}$ using the gate set \{$H$, $P_{N'}$\} admit tower $N$-LimTDD representations for $N \geq 0$ and $N'>0$.
    \item States prepared from $\ket{0}$ using the gate set \{$X$, $CX$, $P_{N'}$, $CP_{N'}$\} admit tower $N$-LimTDD representations for $N \geq 0$ and $N'>0$.
    \item States prepared from $\ket{0}$ using the gate set \{$H$, $CX$, $S$\} admit tower $N$-LimTDD representations for $N \geq 2$.
    \item States prepared from $\ket{0}$ using the gate set \{$H$, $P_{N'}$, $CP_{N'}$\} with at most one $H$ gate per qubit, admit tower $N$-LimTDD representations for $N \geq N'>0$.
\end{enumerate}

The first type of circuit generates no entanglement, while the second type produces no superposition. The third type consists of Clifford circuits. For the fourth type, applying a single $H$ gate per qubit preserves the tower form representation for computational basis states, and phase gates ($P_{N'}$) or controlled phase gates ($CP_{N'}$) maintain this structure without disruption. Note that, before applying an $H$ gate, a qubit remains in $\ket{0}$, as $P_{N'}$ and $CP_{N'}$ can not alter its state.

\begin{figure*}[htbp]
    \centering
    \subfigure[]{
    \includegraphics[width=0.3\textwidth]{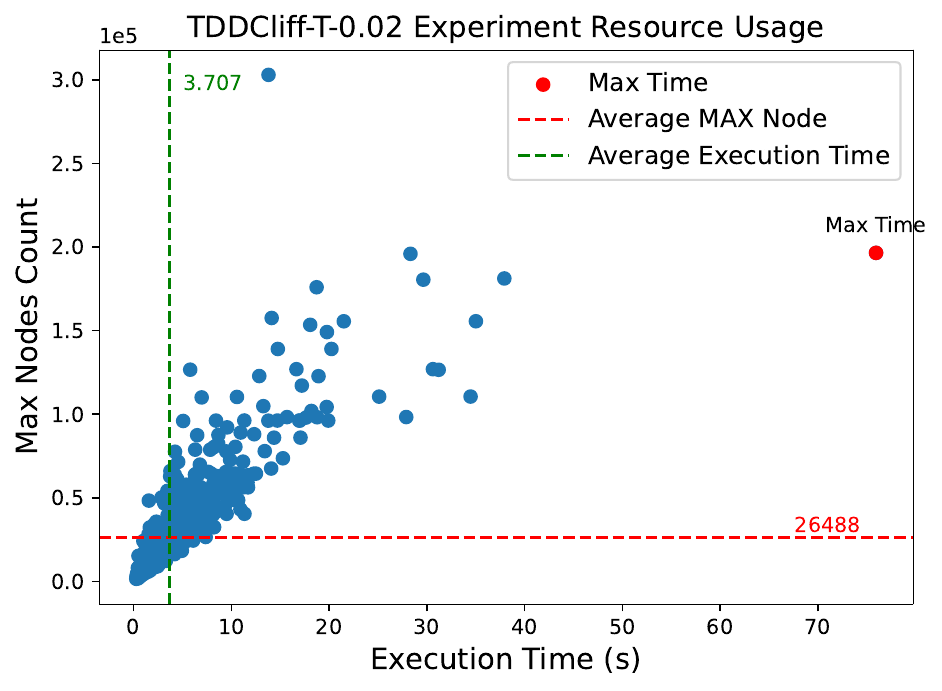}
    }    
    \subfigure[]{
    \includegraphics[width=0.3\textwidth]{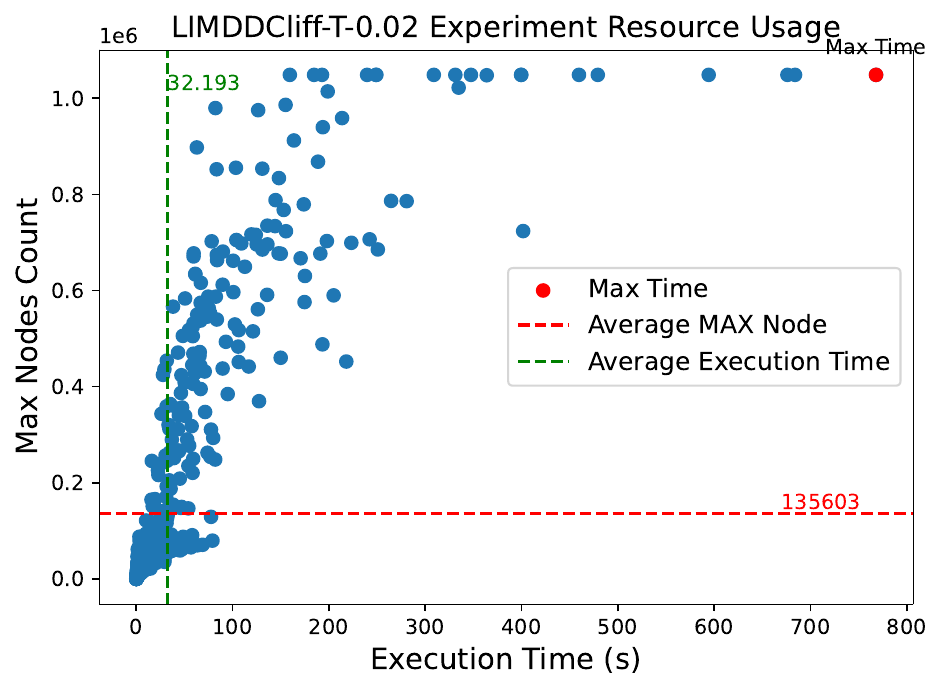}
    }
    \subfigure[]{
    \includegraphics[width=0.3\textwidth]{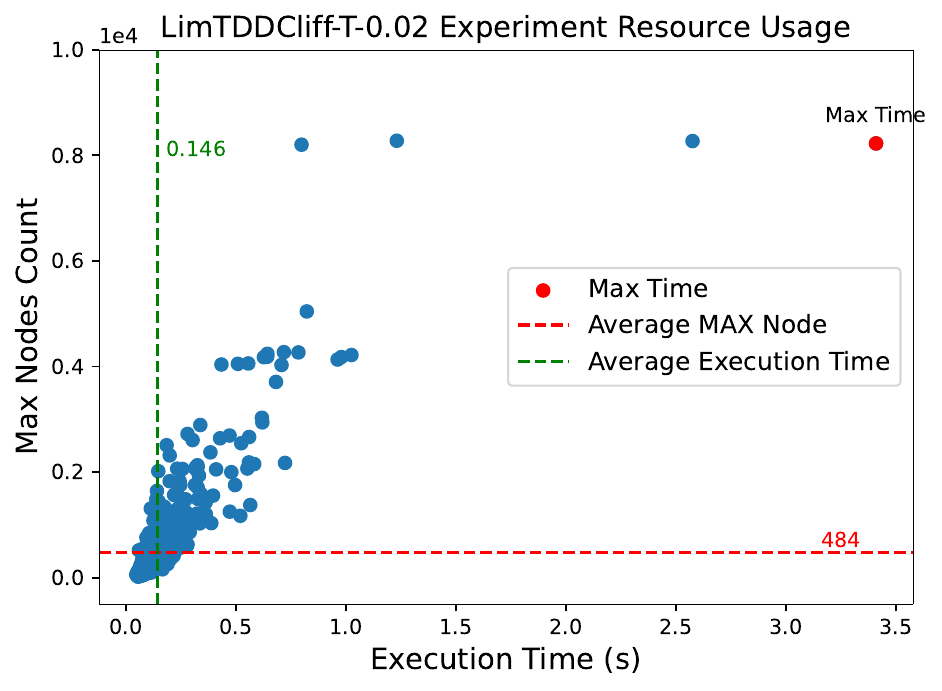}
    }
    \caption{Comparison of time and space efficiency for TDD, LimTDD, and LIMTDD over 1000 random Clifford+T circuits (20 qubits, 600 gates). Average node counts: TDD (\(\approx 26,488\)), LIMDD (\(\approx 135,603\)), LimTDD (\(\approx 484\)).
    The $y$-axes for TDD and LimTDD node counts are scaled using scientific notation (units: $10^5$ for TDD, $10^6$ for LIMDD, and $10^4$ for LimTDD).}
    \label{expe:rand_clifford_t_big}
\end{figure*}

\begin{remark}
\begin{enumerate}
    \item Since LIMDDs are equivalent to 2-LimTDDs for quantum state representations and TDDs correspond to 0-LimTDDs, the first two circuit types admit tower-form representations across all three decision diagrams. The third type (Clifford circuits) requires $N\geq 2$ for tower-form representation in LimTDD. 
    \item \addt{Notably, the fourth type provide examples demonstrating an exponential advantage of LimTDDs over LIMDDs and TDDs. For instance, consider a circuit with $n$ qubits: starting from $\ket{0}^{\otimes n}$, apply an $H$ gate to each qubit, followed by a $CP_{2^{k-1}}$ gate between qubits $k$ and $1$ for each $k \in \{2,\cdots,n\}$. The output state can be represented with only $n+1$ nodes using a $2^n$-LimTDD, but requires $3\cdot 2^{n-2}$ nodes using an LIMDD and $2^n$ nodes using a TDD.}     \addt{Here we continue using the index order $x_n \prec \cdots \prec x_1$. In this order, the $CP_{2^{k-1}}$ gate adds a different phase to the two branches of every $x_k$ node, causing node doubling for TDD when $k>1$. For LIMDD, node doubling starts when $k>2$.}

    \item The theoretical compactness assumes strong canonicity, requiring full stabilizer checks in Algorithm~\ref{alg:LimTDD_norm} (lines 5 and 6). In practice, our implementation omits these checks, which may lead to slightly larger node counts than theoretically predicted. However, experimental data indicate this rarely exceeds TDD/LIMDD sizes (cf. Fig. \ref{expe:rand_clifford_t}, Fig. \ref{expe:rand_clifford_t_big} and Table \ref{tab:functionality}).
\end{enumerate}

\end{remark}

\section{Experiments and Applications}\label{sec:experiments}


We conducted a comprehensive evaluation of LimTDD, TDD, and LIMDD, measuring their efficiency in quantum circuit simulation and functionality construction. Our benchmark suite includes:
\begin{itemize}
\item Standard quantum algorithms: QFT, GHZ, Grover, etc.
\item Random Clifford circuits
\item Random Clifford+T circuits with varying T-gate densities
\end{itemize}
For each method, we recorded total execution time and maximum node count during computation.


All three tools were implemented in C++ and executed on the same platform (Intel(R) Core(TM) i5-13600KF CPU, 32GB RAM,  GCC compiler). The LIMDD implementation is from \addt{\cite{vinkhuijzen2023efficient}} (\url{https://github.com/cda-tum/mqt-limdd}) and the TDD implementation is from (\url{https://github.com/Veriqc/TDD_C}). The C++ version of LimTDD can be found in \url{https://github.com/Veriqc/LimTDD/tree/C}, where a Python version is also provided in the main branch. All three tools use the same index (qubit) order, and all tensors (quantum gates) were processed from left to right according to the circuit, without any additional optimizations. \addt{The current LIMDD implementation~\cite{vinkhuijzen2023limdd,vinkhuijzen2023efficient} lacks support for matrix-matrix product operations, preventing its use in constructing quantum circuit functionality. Thus, we compare only TDDs and LimTDDs for this task. }

\subsection{Simulation with Random Clifford + T Circuits}

\begin{figure*}[htbp]
    \centering
    \subfigure[]{
    \includegraphics[width=0.3\textwidth]{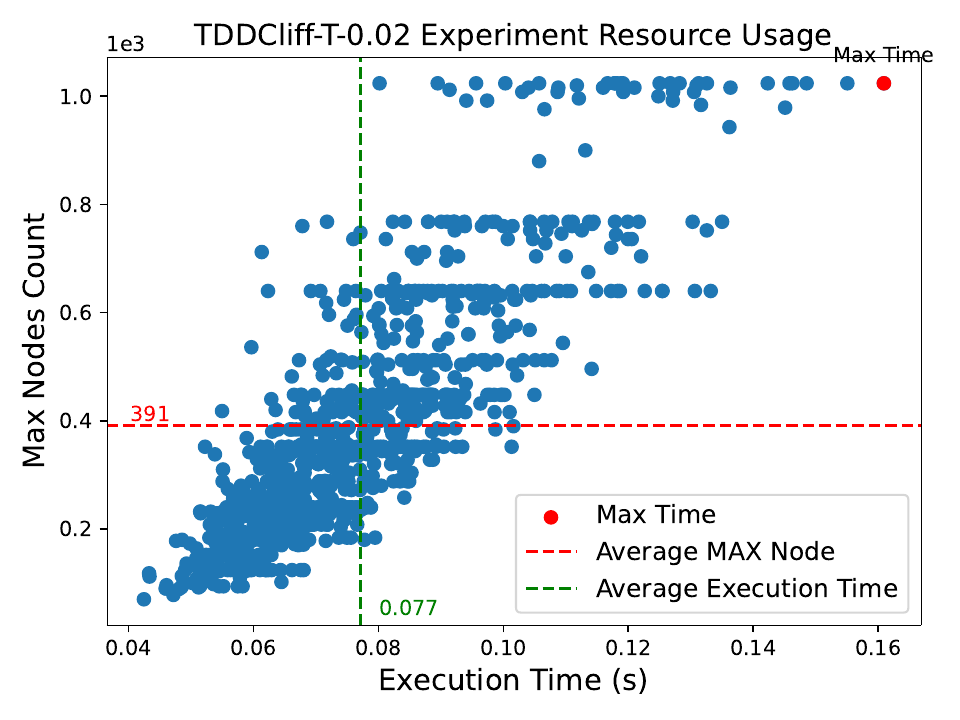}
    }    
    \subfigure[]{
    \includegraphics[width=0.3\textwidth]{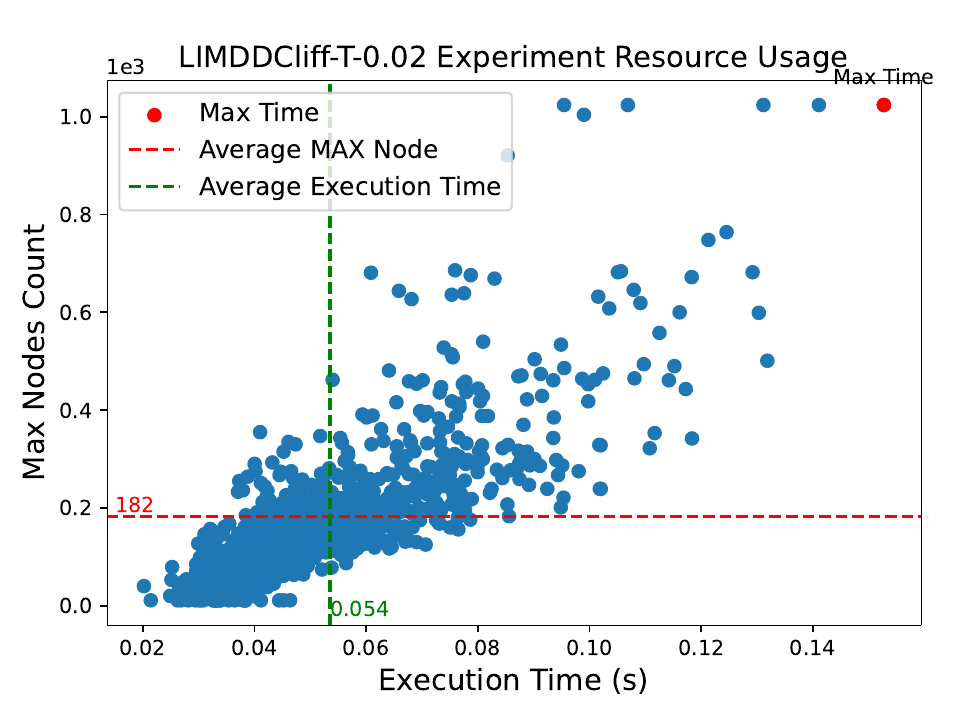}
    }
    \subfigure[]{
    \includegraphics[width=0.3\textwidth]{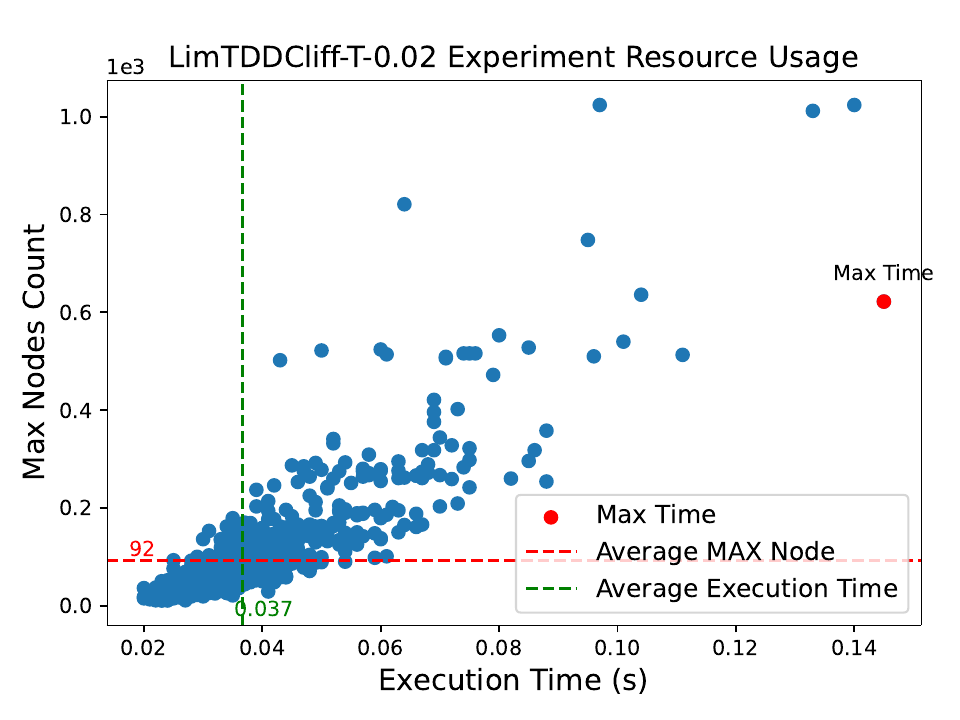}
    }    
    \caption{Comparison of time and space efficiency for TDD, LIMDD, and LimTDD over 1000 random Clifford+T circuits (10 qubits, 400 gates). Average node counts: TDD (\(\approx 391\)), LIMDD (\(\approx 182\)), LimTDD (\(\approx 92\)). The $y$-axes for node counts are scaled using scientific notation (units: $10^3$).
    }
    \label{expe:rand_clifford_t}
\end{figure*}

We first compare the three decision diagrams using random Clifford + T circuits, a universal gate set commonly used in quantum computation. The probability of T-gate occurrence is set to 0.02, and the circuits are simulated with the input state $\ket{0\cdots 0}$. These circuits, 1000 in total, are randomly generated from the gate set $\{X, Y, Z, S, H, CX, T\}$, each consisting of 10 qubits and 400 gates. The corresponding experiment data is shown in Fig.~\ref{expe:rand_clifford_t}. Experimental results show that LimTDDs achieve greater compactness than LIMDDs for these circuits, with TDDs performs the least efficiently, requiring the highest node counts and computation time.

We also conducted experiments on 1000 randomly generated circuits, each with 20 qubits and 600 gates, where the performance trend is even more pronounced. The average runtime for LimTDDs was less than 0.15 seconds, over 20 times faster than TDDs and 200 times faster than LIMDDs. Furthermore, LimTDDs consistently required fewer nodes, with an average maximum of approximately 484 nodes, compared to approximately 26484 nodes for TDDs and 135603 nodes for LIMDDs. Notably, LimTDDs outperformed both methods in efficiency and compactness in 992 out of 1000 experiments. These results underscore LimTDD's superior performance for Clifford + $T$ circuits, highlighting its practical advantages in quantum computation. Besides, LIMDDs performed worse than TDDs in these experiments, possibly due to implementation limitations in the current version of LIMDDs or incomplete adherence to the methodologies described in~\cite{vinkhuijzen2023limdd}.

\subsection{Quantum Circuit Functionality Construction}

We then compare the two decision diagrams, TDDs and LimTDDs, in their ability to construct quantum circuit functionality. For this comparison, we use several standard quantum algorithms, including QFT, GHZ state preparation, and Grover’s algorithm. Functionality representation, which precisely describes quantum circuit behavior, is critical for tasks such as equivalence checking. All circuits are sourced from the MQT benchmark suite~\cite{quetschlich2023mqtbench}.



The comparison results are summarized in Table~\ref{tab:functionality}. In the worst cases, LimTDDs match the node counts of TDDs and require slightly longer computation time due to their more complex internal operations. However, LimTDDs significantly outperform TDDs in node count for most scenarios. For example, for the `qpeexact\_10' circuit, LimTDDs require only 68,010 nodes, whereas TDDs need 466,988 nodes--approximately seven times more. Similar trends are observed for circuits like `ae' and `dj'. In the best cases, such as the `qft' circuit, LimTDDs exhibit exponential improvements in compactness and efficiency over TDDs. For the `qft\_10' circuit, LimTDDs require only 21 nodes, while TDDs require 525,311 nodes. When scaled to 12 qubits, LimTDDs complete the task with just 25 nodes, compared to 8,392,703 nodes for TDDs.

\begin{table}[h]
\centering
\caption{The experiment data for functionality of some common algorithms.}
\label{tab:functionality}
\setlength{\tabcolsep}{0.75mm}{
\scalebox{0.8}{
\begin{tabular}{lcccccc}
\hline
\multicolumn{1}{c}{\multirow{2}{*}{Benchmarks}}   &  & \multicolumn{2}{c}{TDD} &  & \multicolumn{2}{c}{LimTDD} \\ \cline{3-4} \cline{6-7} 
                    &  & \textbf{Time (s)}      & \textbf{Nodes}      &  & \textbf{Time (s)}    & \textbf{Nodes}     \\ \cline{1-1} \cline{3-4} \cline{6-7} 
ae\_10              &  & 1.819       & 174802    &  & 4.651         & 131111     \\
dj\_60              &  & 0.079       & 356        &  & 0.061         & 121         \\
dj\_120             &  & 0.291       & 716        &  & 0.285         & 241         \\
ghz\_60             &  & 0.012       & 355       &  & 0.004         & 121         \\
ghz\_120            &  & 0.033       & 715       &  & 0.007         & 241         \\
graphstate\_30      &  & 0.129       & 34301     &  & 0.004         & 61         \\
graphstate\_60      &  & 75.455      & 3798765   &  & 0.011         & 121         \\
grover-noancilla\_7 &  & 23.187      & 16384     &  & 81.636        & 16384      \\
portfoliovqe\_8     &  & 3.425       & 65536     &  & 5.174         & 65536      \\
qft\_10             &  & 2.416       & 525311    &  & 0.019         & 21         \\
qft\_12             &  & 137.745     & 8392703   &  & 0.074         & 25         \\
qftentangled\_10    &  & 5.840       & 567597    &  & 5.762         & 153158     \\
qnn\_8              &  & 1.938       & 65536     &  & 8.483         & 65536      \\
qpeexact\_10        &  & 0.575       & 26978     &  & 0.138         & 3928       \\
qpeexact\_12        &  & 14.793      & 466988    &  & 2.261         & 68010       \\
qpeinexact\_10      &  & 1.309       & 87334     &  & 0.345         & 15424      \\
qpeinexact\_12      &  & 60.244      & 1379333   &  & 11.398        & 237193      \\
qwalk-noancilla\_7  &  & 3.013       & 15379     &  & 5.606         & 11282      \\
 \hline
\end{tabular}
}
}
\end{table}


\subsection{The Influence of Operator Precision}

\begin{figure}[htbp]
    \centering
    \includegraphics[width=\linewidth]{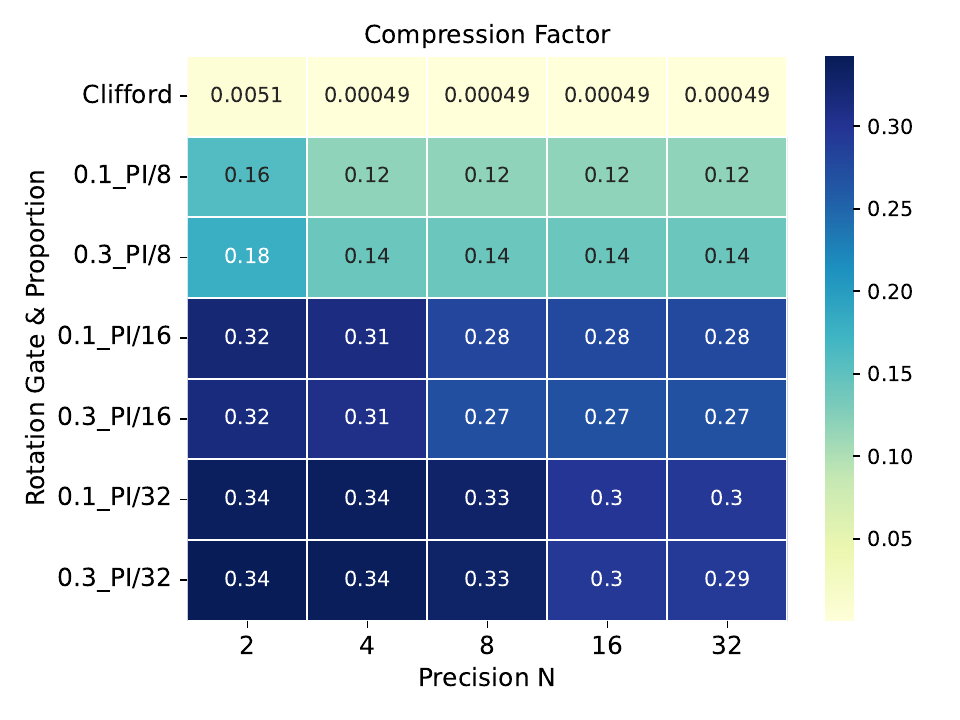}
    \vspace{-1em}
    \caption{Results of simulating random circuits using LimTDD with different precision levels ($N$th root of unity). Rows represent different circuit types (Clifford, and Clifford with various rotation gates and proportions; for example, $0.1\_PI/8$ represents a gate $diag(1,e^{2\pi i/8})$ with proportions 0.1). Data shows the compression factor, defined as the average maximum number of nodes divided by $2^{15}$. We use the gradation from light to dark colors to represent the continuous increase in compression factor.
    }
    \label{tab:rot_p}
\end{figure}

In this subsection, we investigate the impact of precision \(N\) on the compactness of LimTDDs in quantum circuit simulation. We consider four circuit types: Clifford, Clifford+\( T (\text{diag}(1, e^{2\pi i / 8})) \), Clifford+\( \sqrt{T} ( \text{diag}(1, e^{2\pi i / 16})) \), and Clifford+\( \sqrt[4]{T} ( \text{diag}(1, e^{2\pi i / 32}) )\). These circuits are selected due to their progressively decreasing minimum rotation angles, which may require higher precision for accurate simulation. Each circuit consists of 400 gates and 15 qubits, with rotation gate proportions of 0.1 and 0.3. We simulate 200 random circuits for each configuration and record the average maximum number of nodes \((\# \text{Nodes})\) required, along with the compression factor, defined as \(\# \text{Nodes} / 2^{15}\). The results are summarized in Fig.~\ref{tab:rot_p}.


As shown in Fig.~\ref{tab:rot_p}, the compression factor decreases with increasing precision $N$ for all circuit types. However, this trend stops for Clifford+$T$ circuits when $N>4$ and for Clifford+$\sqrt{T}$ circuits when $N>8$. For Clifford+$\sqrt[4]{T}$ circuits, the trend persists up to $N=32$. This shows that higher $N$ enhances LimTDD compactness, underscoring the need for LimTDDs in efficient quantum circuit simulation. However, it also highlights a limitation: beyond a certain $N$, further increases do not improve compactness due to a fundamental limit imposed by the minimum argument of the circuit's matrix elements, necessitating alternative techniques to overcome this constraint.

\subsection{An Example of LimTDD's Exponential Advantage}

\begin{figure}
\centerline{
\begin{tikzcd}[column sep=0.2cm,row sep=0.12cm]
&\qw &\gate{H} &\ctrl{2} &\qw       &\qw  &\qw           &\qw&\qw&\qw\\
&\qw &\gate{H} &\qw      &\ctrl{2}  &\qw  &\qw           &\qw&\qw&\qw\\
&\qw &\qw      &\gate{X} &\qw       &\qw  &\gate[2]{\ \ QFT\ \ } &\qw&\qw&\qw\\
&\qw &\qw      &\qw      &\gate{X}  &\qw  &\qw           &\qw&\qw&\qw\\
\end{tikzcd}

}
    \caption{A quantum circuit for preparing $(I\otimes QFT)\ket{\Psi}$, where $\ket{\Psi} = \frac{1}{\sqrt{2^n}}\sum_{i=0}^{2^n-1}\ket{ii}$.}
    \label{fig:qc_qft}
\end{figure}
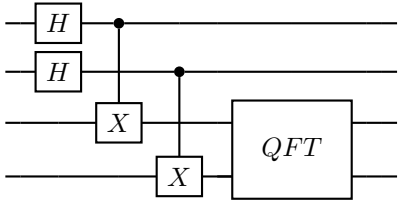

\begin{figure}[h]
    \centering
    \subfigure[]{
    \includegraphics[height=0.3\textwidth]{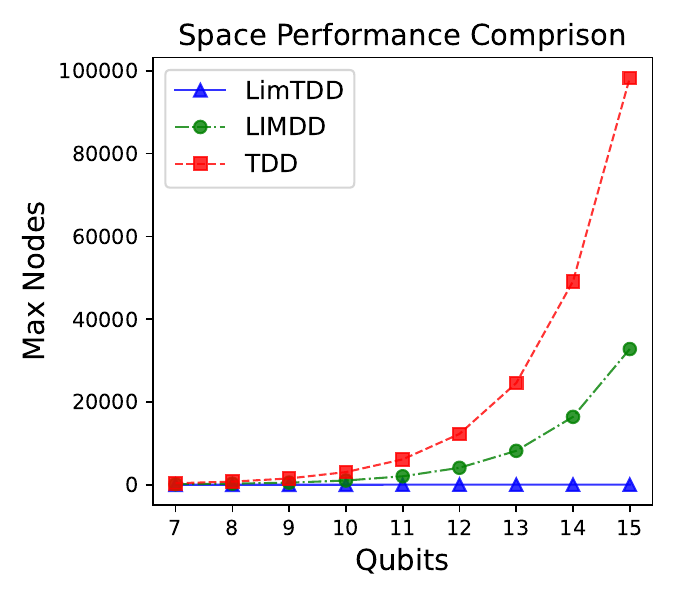}
    }
    \subfigure[]{
    \includegraphics[height=0.3\textwidth]{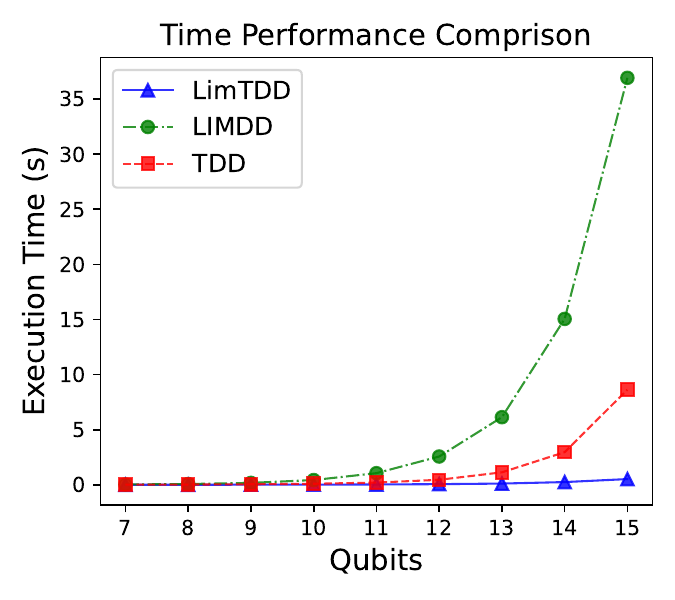}
    }
    \caption{Time and space comparison of TDD, LimDD, and LimTDD in simulating the circuit shown in Fig.~\ref{fig:qc_qft}. LimTDD used a precision ($N$) of $2^{15}=32,768$.
    }
    \label{expe:qft}
\end{figure}

The impact of precision $N$ on LimTDD compactness for various circuit types, combined with the theoretical comparisons in Section~\ref{sec:compare}, motivates us to identify circuits exhibiting an exponential advantage over other decision diagrams, such as TDDs and LIMDDs. This subsection presents such an example, featuring QFT as its core component. A QFT circuit includes Controlled-$\sqrt[k]{T}$ gates for $k$ in $\{ 2^0,2^1,\cdots, 2^{n-3}\}$. Fig.~\ref{expe:qft} presents the experimental results for simulating the circuit depicted in Fig.~\ref{fig:qc_qft}. The circuit applies an $I\otimes QFT$ operation on the maximally entangled state $\ket{\Psi} = \frac{1}{\sqrt{2^n}}\sum_{i=0}^{2^n-1}\ket{ii}$. The results demonstrate that LimTDDs are exponentially more efficient and compact than TDDs and LIMDDs. For $n=15$ qubits, LimTDDs complete the simulation in 0.5 seconds with a maximum node count of 31. In contrast, TDDs requires 8.6 seconds with 98,302 nodes, and LIMDDs take 37 seconds with 32,797 nodes. These results highlight LimTDD's significant advantages in computational efficiency and memory usage.

\section{Conclusion}\label{sec:conclusion}

This paper presented LimTDD, a novel decision diagram that integrates tensor representations with local invertible maps (LIMs) to achieve compact and efficient representations of quantum states and general tensors. By generalizing LIMDD's use of Pauli operators to the more flexible $XP$-stabilizer group, LimTDD extends applicability beyond quantum states to arbitrary tensor networks while maintaining superior compression. Theoretical analysis demonstrates that LimTDDs are at least as compact as TDDs and LIMDDs, with exponential advantages in best-case scenarios, such as circuits involving QFT and controlled-phase gates. Experimental results confirm these improvements, showing significant reductions in node counts and computation times for quantum circuit simulation and functionality computation tasks.

\addt{Based on LimTDD, Hong et al. propose a novel approach for quantum state preparation in \cite{xin2025limtddqsp}. Examples indicate that, in the best-case scenario, the LimTDD-based method can achieve exponential efficiency gains over state-of-the-art algorithms.} Future work will focus on extending LimTDD's operator support to include multi-qubit gates and arbitrary diagonal operations, further enhancing its compression capabilities. Additionally, we plan to investigate applications in broader tensor network computations beyond quantum computing. These advancements will establish LimTDDs as a versatile and powerful tool for efficient representation and manipulation of high-dimensional data in quantum and classical settings.

\bibliographystyle{IEEEtran}
\bibliography{ref}

\section*{Appendix}\label{sec:appendix}


\subsection{Operations of XP-Operators}\label{sec:appxp}

Understanding the group generated by XP-operators requires characterizing the unit operator, multiplication, and inverses of XP-operators. We will need the following notation: for any integer vector $\bold{z} \in \mathbb{Z}^n$, we write $D_N(\bold{z}) := XP_N(\sum_i{\bold{z}[i]}|\bold{0}|{-}\bold{z})$, denoting the antisymmetric operator for $\bold{z}$.

First, we note that  
$$I = XP_N(0|\bold{0}|\bold{0}),\quad  -I=XP_N(N|\bold{0}|\bold{0}).$$

Second, suppose $\bold{u}_1 = (p_1|\bold{x_1}|\bold{z_1})$ and $\bold{u_2} = (p_2|\bold{x_2}|\bold{z_2})$. The multiplication of two XP-operators $XP_N(\bold{u_1})$ and $XP_N(\bold{u_2})$ can then be computed as:
\begin{equation}\label{eq:XPD1}
   XP_N(\bold{u_1})XP_N(\bold{u_2}) = XP_N(\bold{u}_1+\bold{u}_2)D_N(2\bold{x_2}\bold{z_1}),
\end{equation}
where addition and multiplication of vectors are performed component-wise and with proper modulos. That is: 
\begin{eqnarray*}
   (\bold{a}+\bold{b})[j] &=& \bold{a}[j]+\bold{b}[j] \mod{\bold{m}[j]},\\
   (\bold{a}\bold{b})[j] &=& \bold{a}[j]\bold{b}[j] \mod{\bold{m}[j]},
\end{eqnarray*}
where $\bold{m}$ is a $(2n+1)$-dimensional vector such that
\[
\bold{m}[j]=\begin{cases}
		2N, & \text{if $j=0$}\\
           2, & \text{if $1\leq j\leq n$}\\
           N, & \text{if $n+1\leq j\leq 2n$}\\
		 \end{cases}
\]

Lastly, the inverse of the operator $XP_N(p|\bold{x}|\bold{z})$ is given by:
\begin{equation}
\label{eq:XPD2}
XP_N(-p|\bold{x}|-\bold{z})D_N(-2\bold{x}\bold{z}).
\end{equation}
Furthermore, since the $X$-components of $D_N(2\bold{x_2}\bold{z_1})$ and $D_N(-2\bold{x}\bold{z})$ are $\bold{0}$, right multiplication of these operators in \eqref{eq:XPD1} and \eqref{eq:XPD2} only modifies the phase and $Z$-components of the original XP-operator.

\subsection{Calculate $\stab(v)$}\label{sec:stab}

We show how to calculate a generating set for $\stab(v)$ for an internal node $v$, and how to calculate $\min_{g_0 \in \stab(r_{\F_0}),g_1 \in \stab(r_{\F_1})} {g_0^{\dag} \cdot w \cdot g_1}$. 

Suppose $v$ is a node with two successors $v_0$, $v_1$, and the label on the high edge is $h$; then there are two cases. First, if $v_0 \neq v_1$, then only operators of the form $P^z \otimes g$ can be in $\stab(v)$, and $g$ must stabilize $v_0$ and $\omega^{2z} h^{\dagger} g h$ must stabilize $v_1$ for some $z$. Otherwise, operators of the form $P^z \otimes g$ and $XP^z \otimes g$ all can be in $\stab(v)$. More specifically:

(1) If $v_0 \neq v_1$, $\stab(v) = \{ P^z \otimes g | g\in \stab(v_0)\ and\ \omega^{2z} h^{\dagger} g h \in \stab(v_1)\}$;

(2) If $v_0 = v_1$, $\stab(v) = \{ P^z \otimes g | g\in \stab(v_0)\ and\ \omega^{2z} h^{\dagger} g h \in \stab(v_1)\} \cup \{ XP^z \otimes g | hg\in \stab(v_0)\ and\ \omega^{2z} g h \in \stab(v_1) \}$.

For the rank 1 cases, the tensor represented by a normalized node $v$ can only be $[1,r]^\intercal$, where $r \in \mathbb{R}$, and $0 \leq r \leq 1$. Then, there are three cases:

(1) If $r=0$, then $\stab(v) = \langle P \rangle$;

(2) If $r=1$, then $\stab(v) = \langle X \rangle$;

(3) $\stab(v) = \emptyset$, otherwise.

Suppose $G_0$ and $G_1$ are the generating sets of $\stab(v_0)$ and $\stab(v_1)$ ignoring their phase components. To calculate a generating set for $\stab(v)$, we only need to calculate $\langle G_0 \rangle \cap \langle h G_1 h^\dag \rangle$ and then add a suitable phase for each item. Note that the intersection of two groups of XP Stabilizers can be calculated by solving a system of linear equations $S_{X0}^{\vec{a_0}} S_{Z0}^{\vec{b_0}} = S_{X1}^{\vec{a_1}} S_{Z1}^{\vec{b_1}}$ in the canonical generating set form, and the time complexity is polynomial with the length of the operators. The minimum value of $ {g_0^{\dag} \cdot w \cdot g_1}$ for $g_0 \in \stab(r_{\F_0}),g_1 \in \stab(r_{\F_1})$ can also be calculated using their canonical generating set form.

Note that in lines 9 and 10 of Algorithm~\ref{alg:LimTDD_norm}, we are supposed to find the minimal element in the following sets  
\begin{align*}
    \{g_0^{\dag} \cdot w_{\F_0}^\dag \cdot w_{\F_1} \cdot g_1 \ |\ g_0 \in \stab(r_{\F_0}),g_1 \in \stab(r_{\F_1})\} \\
    \{g_1^{\dag} \cdot w_{\F_1}^\dag \cdot w_{\F_0} \cdot g_0 \ |\ g_0 \in \stab(r_{\F_0}),g_1 \in \stab(r_{\F_1})\},
\end{align*}
 which is useful for making the representation canonical. Although it can be done with a time complexity polynomial with $n$, we still think it is a little complicated. In our implementation, we will not obey this routine, and we will suppose $\stab(r_{\F_0}) = \stab(r_{\F_1}) = \{I\}$, then (iii) in Lemma~\ref{lem:norm} will be changed to \[\locnorm\Big(x,\xrightarrow[]{w_1}\Circled{v_1},\xrightarrow[]{w_0}\Circled{v_0}\Big) = \big((X\otimes I) \cdot w, node\big),\] if $w_0^\dag w_1 \neq w_1^\dag w_0$ or $v_0 \neq v_1$; and (iv)  needs $w_0\cdot g_0$ and $w_1\cdot g_1$ to remain in the same order as $w_0$ and $w_1$. This damages the property of canonical to some extent, but it is not serious according to our experimental data. In fact, the complete canonicity is not a very necessary matter for simulation or for verification of quantum circuits since two quantum circuits are equivalent if and only if $tr(UV^\dag)=2^n$ and equivalent up to global phase if and only if $|tr(UV^\dag)|=2^n$. Here, $U$ and $V$ represent the matrix representation of the two circuits, $n$ is the number of qubits, $tr(UV^\dag)$ is a scalar and can be calculated by reversing one of the circuits and connecting the input and output of the two circuits.

\subsection{Proof of Some Theorems and Results}\label{sec:proof}

\textbf{Lemma 1.}\textit{
Suppose $v$ is a node in a LimTDD with form  
\[\Circled{v_{0}}\overset{w_0}{\dashleftarrow}\Circled{v}\xrightarrow[]{w_1}\Circled{v_{1}}.\] 
Let $g_i$ be a $XP$-stabilizer of $\vecc{v_i}$ $(i=0,1)$, and $0\leq k <N$.
Then the following LimTDDs represent the same tensor $\Phi(v)$:
\[\Big( 1 ,\  \Circled{v_{0}}\overset{w_0\cdot g_0}{\dashleftarrow}\Circled{v}\xrightarrow[]{w_1\cdot g_1}\Circled{v_{1}}\Big),\] 
\[\Big(P^ k \otimes (w_0\cdot g_0),\  \Circled{v_{0}}\overset{I}{\dashleftarrow}\Circled{v}\xrightarrow[]{ \omega^{-2k}\cdot g_0^\dagger\cdot w_0^\dagger\cdot w_1\cdot g_1}\Circled{v_{1}}\Big),\] 
\[\Big( X \otimes (w_1\cdot g_1),\  \Circled{v_{1}}\overset{I}{\dashleftarrow}\Circled{v}\xrightarrow[]{g_1^\dagger\cdot w_1^\dagger\cdot w_0\cdot g_0}\Circled{v_{0}}\Big).\]    
}
\begin{proof}
    According to the semantics of LimTDD given in Definition \ref{def:limtdd}, $\Big( 1 ,\  \Circled{v_{0}}\overset{w_0}{\dashleftarrow}\Circled{v}\xrightarrow[]{w_1}\Circled{v_{1}}\Big)$ represents a tensor whose vectorization is $\vecc{v} = [(w_0\vecc{v_0})^\intercal,(w_1\vecc{v_1})^\intercal]^\intercal$.
    
    Similarly, the vectorization of the tensor represented by $\Big( 1 ,\  \Circled{v_{0}}\overset{w_0\cdot g_0}{\dashleftarrow}\Circled{v}\xrightarrow[]{w_1\cdot g_1}\Circled{v_{1}}\Big)$ is $\vecc{v'} = [(w_0g_0\vecc{v_0})^\intercal,(w_1g_1\vecc{v_1})^\intercal]^\intercal$. Since $g_i$ is a stabilizer of $\vecc{v_i}$, that is $g_i\vecc{v_i} = \vecc{v_i}$, $\vecc{v'}=\vecc{v}$.

    The vectorization of the tensor represented by $\Big(P^ k \otimes (w_0\cdot g_0),\  \Circled{v_{0}}\overset{I}{\dashleftarrow}\Circled{v}\xrightarrow[]{ \omega^{-2k}\cdot g_0^\dagger\cdot w_0^\dagger\cdot w_1\cdot g_1}\Circled{v_{1}}\Big)$ is $\vecc{v''} = P^ k \otimes (w_0\cdot g_0)\cdot [(\vecc{v_0})^\intercal,(\omega^{-2k}\cdot g_0^\dagger\cdot w_0^\dagger\cdot w_1\cdot g_1\vecc{v_1})^\intercal]^\intercal = P^ k \otimes I\cdot [(w_0\cdot g_0\vecc{v_0})^\intercal,(\omega^{-2k}\cdot w_1\cdot g_1\vecc{v_1})^\intercal]^\intercal$. Since the function of the $P^k$ gate is to add a phase $\omega^{2k}$ to the second part of the vector, $\vecc{v''}=\vecc{v'}$.

    The vectorization of the tensor represented by $\Big( X \otimes (w_1\cdot g_1),\  \Circled{v_{1}}\overset{I}{\dashleftarrow}\Circled{v}\xrightarrow[]{g_1^\dagger\cdot w_1^\dagger\cdot w_0\cdot g_0}\Circled{v_{0}}\Big)$ is $\vecc{v'''} = X \otimes (w_1\cdot g_1)\cdot [(\vecc{v_1})^\intercal,(g_1^\dagger\cdot w_1^\dagger\cdot w_0\cdot g_0\vecc{v_0})^\intercal]^\intercal = X \otimes I\cdot [(w_1\cdot g_1\vecc{v_1})^\intercal,(w_0\cdot g_0\vecc{v_0})^\intercal]^\intercal$. Since the function of the $X$ gate is to exchange the first and second parts of the vector, $\vecc{v'''}=\vecc{v'}$.
\end{proof}

\textbf{Lemma 2.}\textit{
    Suppose $\locnorm\Big(x,\xrightarrow[]{w_0}\Circled{v_0},\xrightarrow[]{w_1}\Circled{v_1}\Big) = \xrightarrow[]{w}\Circled{v}$. Then we have the following new local normalizations:
    \begin{itemize}
    \item[(i)] $\locnorm\Big(x,\xrightarrow[]{O\cdot w_0}\Circled{v_0},\xrightarrow[]{O\cdot w_1}\Circled{v_1}\Big) = \xrightarrow[]{I\otimes O\cdot w}\Circled{v}$, for any $O \in \mathcal{XP}$; 
    \item[(ii)]  $\locnorm\Big(x,\xrightarrow[]{w_0}\Circled{v_0},\xrightarrow[]{\omega^{2k}w_1}\Circled{v_1}\Big) = \xrightarrow[]{(P^k\otimes I)\cdot w}\Circled{v}$;
    \item[(iii)]  $\locnorm\Big(x,\xrightarrow[]{w_1}\Circled{v_1},\xrightarrow[]{w_0}\Circled{v_0}\Big) =\xrightarrow[]{(X\otimes I) \cdot w}\Circled{v}$, if $v_0\neq v_1$ or $w_0 \neq w_1$;
    \item[(iv)]  $\locnorm\Big(x,\xrightarrow[]{w_0 \cdot g_0}\Circled{v_0},\xrightarrow[]{w_1 \cdot g_1}\Circled{v_1}\Big) = \xrightarrow[]{w}\Circled{v}$, for any $g_0 \in \stab(v_0)$ and $g_1 \in \stab(v_1)$.
    \end{itemize} 
}
\begin{proof}
For convenience, we only prove the case when $v_0< v_1$ and $w_0\neq w_1$. Other cases can be easily extended.

Let $g_0^\star$ and $g_1^\star$, be two operators in $\stab(v_0)$ and $\stab(v_1)$ that minimize $\{g_0'^\dagger\cdot w_0^\dagger\cdot w_1\cdot g_1'\}$ for $g_0'\in \stab(v_0), g_1'\in \stab(v_1)$.
Suppose $w_{temp} =g_0^{\star\dagger}\cdot w_0^\dagger\cdot w_1\cdot g_1^\star =c\cdot e^{i\theta}\omega^{2a} O'$, where $c \in \mathbb{R}$, $0\leq \theta <2\cdot \texttt{ang}(\omega)$, $a\in \mathbb{N}$, and $c(O')=1$. Note that $\texttt{ang}(\omega)$ is the angle of $\omega$, and $c(O')$ is the coefficient of $O'$. Suppose $\omega = e^{2\pi i/N}$, $\texttt{ang}(\omega) = 2\pi/N$. Then  $\locnorm\Big(x,\xrightarrow[]{w_0}\Circled{v_0},\xrightarrow[]{w_1}\Circled{v_1}\Big) = \Big( w ,\  \Circled{v_{0}}\overset{I}{\dashleftarrow}\Circled{v}\xrightarrow[]{c\cdot e^{i\theta} O'}\Circled{v_{1}}\Big)$, where $\idx(v)=x$, and $w = P^a\otimes (w_0\cdot g_0^\star)$.

Then, for $\locnorm\Big(x,\xrightarrow[]{O\cdot w_0}\Circled{v_0},\xrightarrow[]{O\cdot w_1}\Circled{v_1}\Big)$, according to the procedure of Algorithm \ref{alg:LimTDD_norm}, we have $w_{temp}' = g_0^{\star\dagger}\cdot w_0^\dagger\cdot O^\dagger \cdot O \cdot w_1\cdot g_1 = w_{temp}$. Thus, the high-edge weight will still be $c\cdot e^{i\theta} O'$, but the weight of the incoming edge will becomes $P^k\otimes (O\cdot w_0\cdot g_0^\star) = (I\otimes O)\cdot w$.

For $\locnorm\Big(x,\xrightarrow[]{w_0}\Circled{v_0},\xrightarrow[]{\omega^{2k}w_1}\Circled{v_1}\Big)$, $w_{temp}'' = \omega^{2k}\cdot g_0^{\star\dagger}\cdot w_0^\dagger\cdot O^\dagger \cdot O \cdot w_1\cdot g_1 = \omega^{2k} w_{temp}$. Then $\omega^{2k}$ will become a $P^k\otimes I$ operator applied on the incoming edge. Thus, the result will become $\xrightarrow[]{(P^k\otimes I)\cdot w}\Circled{v}$.

For $\locnorm\Big(x,\xrightarrow[]{w_1}\Circled{v_1},\xrightarrow[]{w_0}\Circled{v_0}\Big)$, since $v_0<v_1$, we will exactly exchange the two sub-LimTDDs and enter Algorithm \ref{alg:LimTDD_norm} with $\locnorm\Big(x,\xrightarrow[]{w_0}\Circled{v_0},\xrightarrow[]{w_1}\Circled{v_1}\Big)$ and add a $X\otimes I$ operator on the weight of the incoming edge to indicate that the two successors have been changed. Thus, the result will be $\xrightarrow[]{(X\otimes I) \cdot w}\Circled{v}$.

For the last case, $\locnorm\Big(x,\xrightarrow[]{w_0 \cdot g_0}\Circled{v_0},\xrightarrow[]{w_1 \cdot g_1}\Circled{v_1}\Big)$, since $g_0 \in \stab(v_0)$ and $g_1\in \stab(v_1)$, then $\min_{g_0'\in \stab(v_0), g_1'\in \stab(v_1)} {g_0'}^\dagger\cdot g_0^\dagger\cdot w_0^\dagger\cdot w_1\cdot g_1\cdot g_1' = w_{temp}$. Thus, the result will still be $\xrightarrow[]{w}\Circled{v}$.

\end{proof}

\textbf{Lemma 3.}\textit{
Let $\F$ be the LimTDD of tensor $\phi_{x_n\cdots x_1}$. For $c\in \{0,1\}$, define $\F_{x_i=c} = \texttt{Slicing}(\F,x_i,c)$. Then, $\F_{x_i=c}$ is the LimTDD representation of the tensor $\phi_{x_i=c}$.
}

\begin{proof}
We only prove the case when $x_i = x_n$, i.e., the top index of the LimTDD, and $c=0$. Suppose $\F$ is in the form $\Big( \lambda\cdot  X^{b_n}\cdot P^{z_n} \otimes w,\  \Circled{v_{0}}\overset{w_0}{\dashleftarrow}\Circled{v}\xrightarrow[]{w_1}\Circled{v_{1}}\Big)$, representing the tensor $\phi_{x_n\cdots x_1}$. Here, $b_n \in {0,1}$, $z_n \in \{0,\cdots N-1\}$, where $N$ is the precision of the XP-operators, and $w$ is an operator with $c(w)=1$. Then the vectorization of the tensor $\phi$ is $\vecc{\phi} = \lambda\cdot  X^{b_n}\cdot P^{z_n} \otimes w \cdot [(w_0\vecc{v_0})^\intercal, (w_1\vecc{v_1})^\intercal]^\intercal = X^{b_n} \cdot [(\lambda\cdot  w\cdot w_0\vecc{v_0})^\intercal, (\lambda \omega^{2z_n}\cdot w\cdot w_1\vecc{v_1})^\intercal]^\intercal$. Thus, if $b_n = 0$, $\vecc{\phi_{x_i=0}} = (\lambda\cdot  w\cdot w_0\vecc{v_0})^\intercal$; else $b_n =1$ and $\vecc{\phi_{x_i=0}} = (\lambda\cdot  \omega^{2z_n}\cdot w\cdot w_1\vecc{v_1})^\intercal$.

Suppose $b_n = 0$, then the LimTDD obtained through the Slicing procedure given in Algorithm \ref{alg:LimTDD_Slicing} is $\xrightarrow[]{\lambda\cdot  w\cdot w_0}\Circled{v_0}$. If $b_n = 1$, then the LimTDD obtained is $\xrightarrow[]{\lambda \cdot \omega^{2z_n}\cdot w\cdot w_1}\Circled{v_1}$. Both cases represent the tensor $\phi_{x_i=0}$.

\end{proof}

\textbf{Lemma 4.}\textit{
Let $\F$ be the LimTDD representation of tensor $\phi_{x_n\cdots x_1}$. Then 
    $\locnorm(x_n,\F_{x_n=0},\F_{x_n=1}) = \F$.
}

\begin{proof}

Suppose $\F$ is in the form $\Big( \lambda\cdot  X^{b_n}\cdot P^{z_n} \otimes w,\  \Circled{v_{0}}\overset{w_0}{\dashleftarrow}\Circled{v}\xrightarrow[]{w_1}\Circled{v_{1}}\Big)$, representing the tensor $\phi_{x_n\cdots x_1}$. Here, we only consider the case that $b_n=0$, $v_0 < v_1$. Then, according to the Slicing procedure given in Algorithm \ref{alg:LimTDD_Slicing}, $\F_{x_n=0} = \xrightarrow[]{\lambda\cdot  w\cdot w_0}\Circled{v_0}$ and $\F_{x_n=1} = \xrightarrow[]{\lambda\cdot  \omega^{2z_n}\cdot w\cdot w_1}\Circled{v_1}$. Then, according to Lemma \ref{lem:norm} (i) and (ii), $\locnorm(x_n,\F_{x_n=0},\F_{x_n=1}) = \xrightarrow[]{\lambda \cdot P^{z_n}\otimes w}\Circled{v} = \F$.
    
\end{proof}

\textbf{Theorem 1.}\textit{
Suppose $\phi_{x_n\cdots x_1}$ and $\gamma_{x_n\cdots x_1}$ are two tensors with the same index set. Let $\F$ and $\G$ be the $N$-LimTDD representations of $\phi_{x_n\cdots x_1}$ and $\gamma_{x_n\cdots x_1}$ with the same index order. The following two conditions are equivalent:
\begin{enumerate}
    \item $\phi_{x_n\cdots x_1}$ and $\gamma_{x_n\cdots x_1}$ are $\mathcal{XP_N}$-isomorphic.
\item $r_\F =r_\G$ and $w_\F = O \cdot w_\G \cdot g$ for some $O\in \mathcal{XP_N}$ and some  $g \in \stab(r_\G)$.
\end{enumerate} 
}

\begin{proof}
$2\Rightarrow 1$ follows directly from the definition. We prove $1\Rightarrow 2$ by induction on the tensor rank. 

For rank 0 tensors, the claim holds trivially. Now, assume it holds for rank $n-1$, and consider the rank-$n$ case. Given that  $\vecc{\phi_{x_n\cdots x_1}} = O \vecc{\gamma_{x_n\cdots x_1}}$ with $O= X^{b_n}P^{z_n}\otimes w$, we show $r_\F=r_\G$, and  $w_\F = O \cdot w_\G \cdot g$ for some $g \in \stab(\Phi(r_\G))$.

Suppose $\G_{x_n=0} = \xrightarrow[]{w_0}\Circled{v_0}$ and $\G_{x_n=1} = \xrightarrow[]{w_1}\Circled{v_1}$. Then $\G = \locnorm\Big(x_n,\xrightarrow[]{w_0}\Circled{v_0},\xrightarrow[]{w_1}\Circled{v_1}\Big) = (w_\G, r_\G)$, because $\G$ is normalized.

According to the value of $b_n$, there are two cases to examine:

\textbf{Case 1}: If $b_n = 0$, then the components satisfy:
$$\Phi(\F_{x_n=0}) = \phi_{x_n=0} = w \gamma_{x_n=0} = w \Phi(\G_{x_n=0}),$$ 
$$\Phi(\F_{x_n=1}) = \phi_{x_n=1} = \omega^{2z_n}\cdot w \gamma_{x_n=1} = \omega^{2z_n}\cdot w \Phi(\G_{x_n=1}).$$ 
By the induction hypothesis, there exist $g_0 \in \stab(v_0)$ and $g_1 \in \stab(v_1)$ such that:
\[\F_{x_n=0} = \xrightarrow[]{w\cdot w_0 \cdot g_0}\Circled{v_0}\ \text{and}\ \F_{x_n=1} = \xrightarrow[]{\omega^{2z_n} w\cdot w_1\cdot g_1}\Circled{v_1}.\]
Because
\[\F = \locnorm\Big(x_n,\xrightarrow[]{w\cdot w_0 \cdot g_0}\Circled{v_0},\xrightarrow[]{ \omega^{2z_n} w\cdot w_1 \cdot g_1}\Circled{v_1}\Big),\]  
applying Lemma~\ref{lem:norm} in the order (i), (ii), and (iv), we conclude $\F \cong \locnorm\Big(x_n,\xrightarrow[]{w_0}\Circled{v_0},\xrightarrow[]{w_1}\Circled{v_1}\Big) = \G$. 

\textbf{Case 2}: If $b_n=1$, then the components satisfy: 
$$\Phi(\F_{x_n=0}) = \phi_{x_n=0} = \omega^{2z_n}\cdot w \gamma_{x_n=1} = \omega^{2z_n}\cdot w \Phi(\G_{x_n=1}),$$
$$\Phi(\F_{x_n=1}) = \phi_{x_n=1} = w \gamma_{x_n=0} = w \Phi(\G_{x_n=0}).$$
There exist $g_0 \in \stab(v_0)$ and $g_1 \in \stab(v_1)$ such that:
\[\F_{x_n=0} = \xrightarrow[]{\omega^{2z_n} w\cdot w_1 \cdot g_1}\Circled{v_1}\ \text{and}\ \F_{x_n=1} = \xrightarrow[]{w \cdot w_0\cdot g_0}\Circled{v_0}.\]
Because
\[\F = \locnorm\Big(x_n,\xrightarrow[]{\omega^{2z_n} w\cdot w_1 \cdot g_1}\Circled{v_1},\xrightarrow[]{w \cdot w_0\cdot g_0}\Circled{v_0}\Big),\] 
applying Lemma~\ref{lem:norm} in the order (i), (ii), (iii), and (iv), we conclude
\[\F \cong  \locnorm\Big(x_n,\xrightarrow[]{w_0}\Circled{v_0},\xrightarrow[]{w_1}\Circled{v_1}\Big)=G.\]

This means that, for both cases, $r_\F =r_\G$. Since $w_\F \cdot \Phi(r_\F) =\vecc{\phi_{x_n\cdots x_1}} = O\cdot \vecc{\gamma_{x_n\cdots x_1}} =O\cdot w_\G \cdot \Phi(r_\G)$, we have $w_\G^\dag \cdot O^\dag \cdot w_\F \cdot \Phi(r_\G) = \Phi(r_\G)$. Let $g=w_\G^\dag \cdot O^\dag \cdot w_\F$. Then $g \in \stab(\Phi(r_\G))$ and $w_\F = O \cdot w_\G \cdot g$.
\end{proof}

\textbf{Corollary 1.}\textit{
For any tensor $\phi$ and any $N\geq 0$, let $\F$ be its $(N+1)$-LimTDD representation and $\G$ its $N$-LimTDD representation. Then, $size(\F)\leq size(\G)$, where $size(\F)$ and $size(\G)$ denote the number of nodes in $\F$ and $\G$, respectively.
}

\begin{proof}
This is a direct corollary of Theorem \ref{thm:iso_nodes}
and the fact that $\mathcal{XP_N} \subset \mathcal{XP}_{\mathcal{N}+1}$.
    
\end{proof}

\textbf{Theorem 2.}(LimTDD vs. TDD \& LIMDD)\textit{
\begin{enumerate}
    \item Let $\F$ be the LimTDD representation and $\G$ the TDD representation of the same tensor $\phi_{x_n\cdots x_1}$ with the same index order. Then, $size(\F)\leq size(\G)$.
    \item Let $\F$ be the N-LimTDD ($N\geq 2$) representation and $\G$ the LIMDD representation of the same quantum state $\ket{\phi_{x_n\cdots x_1}}$ with the same index (qubit) order. Then, $size(\F)\leq size(\G)$.
\end{enumerate}
}

\begin{proof}
    Denote $\mathcal{XP}_0 = \{c\cdot I, c\in \mathbb{C}\}$. Since TDD only allows two tensors that differ only by a complex number to share the same node. Thus, TDD can be seen as 0-LimTDD, and when representing quantum states LIMDD can be seen as 2-LimTDD. Thus, this is a direct corollary of Corollary \ref{cor:size}.
\end{proof}

\end{document}